\documentclass{LMCS}

\usepackage{array}
\usepackage{graphicx}
\usepackage{epic, eepic}
\usepackage{amssymb,amsmath}
\usepackage{latexsym}
\usepackage{subfigure}
\usepackage[usenames]{color}
\usepackage{multirow}
\usepackage{gastex}
\usepackage{subfigure}
\usepackage{enumerate,hyperref}
\usepackage{array,tikz}
\usetikzlibrary{arrows,automata,chains,matrix,positioning,scopes,calc,shapes}

\def\mlap#1{\hbox to0 pt{\hss#1\hss}}
\tikzset{rdbox/.style={
  rounded rectangle,
  minimum size=6mm,
  thick,draw=black,
  font=\sffamily}}

\newcounter{compressEnum}
\renewcommand{\thecompressEnum}{$\roman{compressEnum}$}
\newenvironment{compressEnum}
{\setcounter{compressEnum}{0}}% \begin{paragraph}}
{}%\end{paragraph}}
\newcommand{\itCompress}{\stepcounter{compressEnum}{(\thecompressEnum) }}

\def\endof{%
  \leavevmode
  \parfillskip=0pt%
  \widowpenalty=10000%
  \displaywidowpenalty=10000%
  \finalhyphendemerits=0%
  \unskip\nobreak\null\hfil\penalty50\hskip2em\null\hfill%
}
\def\eodsymbol{\ensuremath\square}

\def\origEOD{\nobreak\leavevmode\endof\eodsymbol\par}
\def\EOD{\origEOD\global\def\EOD{}}

\def\abs#1{\ensuremath{\lvert #1\rvert}}

\newcommand{\nat}{\mathbb N} 
\newcommand{\rat}{{\mathbb Q}}

\newcommand{\posreal}{{\mathbb R}^{\geq 0}}
\newcommand{\sposreal}{{\mathbb R}^{> 0}}
\newcommand{\real}{{\mathbb R}}

\newcommand{\tuple}[1]{\langle #1 \rangle}

\newcommand{\C}{\mathcal{C}}

\newcommand{\Last}{{\sf Last}}

\newcommand{\La}{{\sc Lavg}}
\newcommand{\Li}{{\sc Linf}}
\newcommand{\Ls}{{\sc Lsup}}
\newcommand{\Di}{{\sc Disc}}

\newcommand{\nmax}{{\sc NSup}}
\newcommand{\nDi}{{\sc NDisc}}
\newcommand{\nla}{{\sc NLavg}}
\newcommand{\nli}{{\sc NLinf}}
\newcommand{\nls}{{\sc NLsup}}
\newcommand{\nbw}{{\sc NBW}}

\newcommand{\dmax}{{\sc DSup}}
\newcommand{\ddi}{{\sc DDisc}}
\newcommand{\dla}{{\sc DLavg}}
\newcommand{\dli}{{\sc DLinf}}
\newcommand{\dls}{{\sc DLsup}}
\newcommand{\dbw}{{\sc DBW}}

\newcommand{\slopefrac}[2]{\leavevmode\kern.1em
  \raise .5ex\hbox{\the\scriptfont0 #1}\kern-.1em
  /\kern-.15em\lower .25ex\hbox{\the\scriptfont0 #2}}

\newcommand{\nDli}{{\sc \slopefrac{N}{D}Linf}}
\newcommand{\nDmax}{{\sc \slopefrac{N}{D}Sup}}
\newcommand{\nD}{{\sc \slopefrac{N}{D}}}

\newcommand{\Val}{\mathsf{Val}}
\newcommand{\Maxf}{\mathsf{Max}}
\newcommand{\Max}{\mathsf{Sup}}
\newcommand{\LimSup}{\mathsf{LimSup}}
\newcommand{\LimInf}{\mathsf{LimInf}}
\newcommand{\LimAvg}{\mathsf{LimAvg}}
\newcommand{\Disc}{\mathsf{Disc}}
\newcommand{\Avg}{\mathsf{Avg}}
\newcommand{\Sum}{\mathsf{Sum}}

\def\set#1{\ensuremath{\{#1\}}}
\newcommand{\ok}{\checkmark}
\newcommand{\ko}{$\times$}
\newcommand{\weight}{\gamma}

\newcommand{\reward}{v}

\DeclareMathOperator{\op}{op}
\makeatletter

\begingroup \catcode `|=0 \catcode `[= 1
\catcode`]=2 \catcode `\{=12 \catcode `\}=12
\catcode`\\=12 |gdef|@xcomment#1\end{comment}[|end[comment]]
|endgroup

\def\@comment{\let\do\@makeother \dospecials\catcode`\^^M=10\def\par{}}

\def\begincomment{\@comment\@xcomment}

\makeatother

\newenvironment{comment}{\begincomment}{}

\def\doi{6 (3:10) 2010}
\lmcsheading%
{\doi}
{1--23}
{}
{}
{Dec.~15, 2009}
{Aug.~30, 2010}
{}

\begin{document}

\title{Expressiveness and Closure Properties for Quantitative Languages\rsuper*}

\author[K.~Chatterjee]{Krishnendu Chatterjee\rsuper a} 
\address{{\lsuper{a,c}}IST Austria (Institute of Science and Technology Austria)}
\email{\{Krishnendu.Chatterjee, tah\}@ist.ac.at}
\author[L.~Doyen]{Laurent Doyen\rsuper b}
\address{{\lsuper b}LSV, ENS Cachan \& CNRS, France}
\email{doyen@lsv.ens-cachan.fr}
\author[T.~A.~Henzinger]{Thomas A.~Henzinger\rsuper c}
\address{{\lsuper c}EPFL, Lausanne, Switzerland}

\thanks{This 
research was supported in part by the Swiss National Science Foundation under the Indo-Swiss Joint Research Programme,
by the European Network of Excellence on Embedded Systems Design (ArtistDesign), by the European projects Combest, Quasimodo, and Gasics, %
%({\tt http://www.quasimodo.aau.dk/}), 
%({\tt http://www.ulb.ac.be/di/gasics/}), 
by the PAI program Moves funded by the Belgian Federal Government, % ({\tt http://moves.ulb.ac.be}), 
and by the CFV (Federated Center in Verification) funded by the F.R.S.-FNRS. % ({\tt http://www.ulb.ac.be/di/ssd/cfv/}).
}
\titlecomment{{\lsuper*}A preliminary version of this paper appeared in the
  Proceedings of the 24th Annual Symposium on Logic in Computer
  Science (LICS), IEEE Computer Society Press, 2009.}


\begin{abstract}
Weighted automata are nondeterministic automata with numerical weights on
transitions. They can define quantitative languages~$L$ that
assign to each word~$w$ a real number~$L(w)$.
In the case of infinite words, the value of a run is naturally computed as
the maximum, limsup, liminf, limit-average, or discounted-sum of the
transition weights. The value of a word $w$ is the supremum of the values of 
the runs over $w$.
We study expressiveness and closure questions about these quantitative
languages.

We first show that the set of words with value greater than a threshold
can be non-$\omega$-regular for deterministic limit-average and
discounted-sum
automata, while this set is always $\omega$-regular when the threshold
is isolated (i.e., some neighborhood around the threshold contains 
no word).
In the latter case, we prove that the $\omega$-regular language is robust
against small perturbations of the transition weights.

We next consider automata with transition weights $0$ or $1$ and show that they
are as expressive as general weighted automata in the limit-average case, but not
in the discounted-sum case.

Third, for quantitative languages $L_1$ and~$L_2$, we consider the operations
$\max(L_1,L_2)$, $\min(L_1,L_2)$, and $1-L_1$, which generalize
the boolean operations on languages, as well as the sum $L_1 + L_2$.
We establish the closure properties of all classes
of quantitative languages with respect to these four operations.
\end{abstract}

\maketitle\vfill

\section{Introduction}
A boolean language $L$ can be viewed as a function that assigns to each word $w$ a boolean value,
namely, $L(w) = 1$ if the word $w$ belongs to the language, and $L(w) = 0$ otherwise. 
Boolean languages model the computations of reactive programs. The verification
problem ``does the program~$A$ satisfy the specification~$B$?'' 
then reduces to the language-inclusion problem ``is $L_A \subseteq L_B$?'',
or equivalently, ``is $L_{A}(w)\leq L_{B}(w)$ for all words~$w$?'',
where $L_A$ represents the behaviors of the program, and $L_B$ contains 
all behaviors allowed by the specification. When boolean languages are defined
by finite automata, this framework
is called the \emph{automata-theoretic approach} to model-checking~\cite{VardiW86}.

In a natural generalization of this framework, a cost function
assigns to each word a real number instead of a boolean value. 
For instance, the value of a word (or behavior) can be interpreted as the amount of some resource 
(e.g., memory consumption, or power consumption) that the program needs to produce it,
and a specification may assign a maximal amount of available resource to each 
behavior, or bound the long-run average available use of the resource. 

Weighted automata over semirings (i.e., finite automata with transition weights 
in a semiring structure) have been used to define cost functions, 
called formal power series for finite words~\cite{Wautomata,KuichS86}
and $\omega$-series for infinite words~\cite{CulikK94,DrosteK03,EsikK04}.
In~\cite{CDH08}, we study new classes of cost functions using 
operations over rational numbers that do not form a semiring. 
We call them \emph{quantitative languages}.
We set the value of a (finite or infinite) word $w$ as the supremum value of all runs over~$w$
(if the automaton is nondeterministic, then there may be many runs over~$w$), 
and the value of a run $r$ is a function of the (finite or infinite) sequence 
of weights that appear along~$r$. We consider several 
functions, such as $\Maxf$
%
% Laurent: this footnote is made precise after the definition of Weighted automata. 
%
% \footnote{It should be noted that 
% $\Maxf$ generalizes finite-word acceptance condition, while
% $\LimSup$ and $\LimInf$ generalize B\"uchi and coB\"uchi 
% acceptance conditions.\label{ftn:boolean}} 
and $\Sum$ of weights for finite runs, 
and $\Max$, $\LimSup$, $\LimInf$, % \refmark{ftn:boolean} 
limit-average, and discounted sum of weights for infinite runs.
For example, peak power consumption can be modeled as the maximum of a sequence 
of weights representing power usage; energy use can be modeled as the sum;
average response time as the limit-average \cite{CCHK+05,CAHS03}.
Quantitative languages can also be used to specify and verify reliability 
requirements: if a special symbol $\bot$ is used to denote failure and has 
weight~$1$, while the other symbols have weight $0$, one can use a limit-average 
automaton to specify a bound on the rate of failure in the long run~\cite{CGHIKPS08}.
The discounted sum can be used to specify that failures happening
later are less important than those happening soon~\cite{AHM03}.

The \emph{quantitative language-inclusion problem} ``given two automata~$A$ and~$B$, 
is $L_{A}(w)\leq L_{B}(w)$ for all words~$w$?'' can then be used to check, say, if for each behavior, 
the peak power used by the system lies below the bound given by the specification; 
or if for each behavior, the long-run average response time of the system
lies below the specified average response requirements.
In~\cite{CDH08}, we showed that the quantitative language-inclusion problem
is PSPACE-complete for $\Max$-, $\LimSup$-, and $\LimInf$-automata, while
the decidability is unknown for (nondeterministic) limit-average and discounted-sum automata.
We also compared the expressive power of the different classes of quantitative languages
and showed that %, surprisingly,   ??
nondeterministic automata are strictly more expressive than deterministic automata in the 
limit-average and discounted-sum cases. 
%This is not surprising as the quantitative language-inclusion problem
%is solvable %(in polynomial time) when the automata are deterministic.

%Quantitative languages provide a framework to specify quantitative 
%properties of interest: this is illustrated through examples in  
%Section~\ref{sec-examples}.
In this paper, we investigate alternative ways of comparing the \emph{expressive power}
of weighted automata. First, we consider the cut-point languages of weighted
automata, a notion borrowed from the theory of probabilistic automata~\cite{Rabin63}.
Given a threshold $\eta \in \real$, the cut-point language of a quantitative
language $L$ is the set of all words $w$ with value $L(w) \geq \eta$, thus
a boolean language.
We show that deterministic limit-average and 
discounted-sum automata can define cut-point languages that are not 
$\omega$-regular. 
Note that there also exist $\omega$-regular languages that cannot be expressed
as a cut-point language of a limit-average or discounted-sum automaton~\cite{CDH08}.
Then, we consider the special case where the threshold~$\eta$ is isolated, meaning
that there is no word with a value in the neighborhood of~$\eta$. We argue that 
isolated cut-point languages are robust, by showing that they
remain unchanged under small perturbations of the transition weights. 
Furthermore, we show that every discounted-sum automaton with isolated cut-point defines
an $\omega$-regular language, and the same holds for deterministic limit-average
automata. This question is open for nondeterministic limit-average automata.
Finally, we consider a boolean counterpart of limit-average and discounted-sum 
automata in which all transitions have weight~$0$ or~$1$.
%%%\cbstart
Of special interest is a proof that for every limit-average automaton with rational weights in 
the interval $[0,1]$ there is an equivalent limit-average automaton with boolean weights.
Therefore, the restriction to boolean weights does not change the class of quantitative 
languages definable by limit-average automata; on the other hand, we show that it reduces 
the expressive power of discounted-sum automata. 
%%%\cbend

In the second part of this paper, we study the \emph{closure properties} of quantitative languages.
It is natural and convenient to decompose a specification or a design 
into several components, and to apply composition operators
to obtain a complete specification. 
% It is natural and convenient to develop parts of a specification in isolation 
% and then compose them to obtain a complete specification.
% This presents a clear motivation to study the closure properties. 
We consider a natural generalization of the classical operations
of union, intersection, and complement of boolean languages. We define the \emph{maximum}, 
\emph{minimum}, and \emph{sum} of two quantitative languages $L_1$ and $L_2$ as the quantitative language
that assigns $\max(L_1(w),L_2(w))$, $\min(L_1(w),L_2(w))$, and $L_1(w) + L_2(w)$
to each word $w$. 
The \emph{complement}~$L^c$ of a quantitative language~$L$ is defined by $L^c(w) = 1-L(w)$
for all words $w$.\footnote{One can define $L^c(w) = k-L(w)$ for any rational constant $k$ without changing 
the results of this paper.}   % (see Section~\ref{sec:closure-properties}).}
%Maximum and minimum generalize union and intersection respectively, while sum has
%no boolean counterpart. 
The sum is a natural way of composing two automata
if the weights represent costs (e.g., energy consumption). 
We give other examples in Section~\ref{sec:definitions} to illustrate 
the composition operators and the use of quantitative languages 
as a specification framework.

\begin{table}
\begin{center}
\begin{tabular}{|l|*{4}{c|}}
\hline
       & max. & min. & comp. & sum \\
\hline
$\Maxf$   &  \ok & \ok  & \ko   & \ok \\
\hline
$\Last$  &  \ok & \ok  & \ok   & \ok \\
\hline
Det. $\Sum$   &  \ko & \ok  & \ok   & \ok \\
\hline
Nondet.  $\Sum$   &  \ok & \ko  & \ko   & \ok \\
\hline
\multicolumn{5}{c}{(a) Finite words}
\end{tabular}\hfill
\begin{tabular}{|l|*{4}{c|}}
\hline
       & max. & min. & comp. & sum \\
\hline
\nDmax &  \ok & \ok  & \ko   & \ok \\
\hline
\nDli  &  \ok & \ok  & \ko   & \ok \\
\hline
\dls   &  \ok & \ok  & \ko   & \ok \\
\hline
\nls   &  \ok & \ok  & \ok   & \ok \\
\hline
\dla   &  \ko & \ko  & \ko   & \ko \\
\hline
\nla   &  \ok & \ko  & \ko   & \ko \\
\hline
\ddi   &  \ko & \ko  & \ok   & \ok \\
\hline
\nDi   &  \ok & \ko  & \ko   & \ok \\
\hline
\multicolumn{5}{c}{(b) Infinite words}
\end{tabular}\hfill
\end{center}
\caption{Closure properties. The meaning of the acronyms is described on p.\pageref{page:notation}.\label{tab:closure-properties}}
\end{table}

We give a complete picture of the closure properties of 
the various classes of quantitative languages (over finite and infinite words) under maximum,
minimum, complement and sum (see Table~\ref{tab:closure-properties}).
For instance, (non)deterministic limit-average automata are not closed under sum and complement,
while nondeterministic discounted-sum automata are closed under sum but not under complement.
All other classes of weighted automata are closed under sum.
%%%\cbstart
For infinite words, the closure properties of $\Max$-, $\LimSup$-, and $\LimInf$-automata
are obtained as a direct extension of the results for boolean finite automata,
while for limit-average and discounted-sum automata, the proofs require the analysis of the structure
of the automata cycles and properties of the solutions of polynomials with rational coefficients.
%%%\cbend
Note that the quantitative language-inclusion problem 
``is $L_{A}(w)\leq L_{B}(w)$  for all words~$w$?'' reduces to 
closure under sum and complement, because it is equivalent 
to the question of the non-existence of a word $w$ such that $L_{A}(w) + L^c_{B}(w) > 1$,
an \emph{emptiness} question which is decidable for all classes of quantitative languages~\cite{CDH08}.
Also note that deterministic limit-average and discounted-sum automata are not 
closed under maximum, which implies % the result of~\cite{CDH08} 
that nondeterministic automata are strictly more expressive in these cases 
(because the maximum can be obtained by an initial nondeterministic choice). 
\smallskip\noindent{\it Related work.} Functions such as limit-average (or mean-payoff) and discounted 
sum have been studied extensively in the branching-time context of game theory~\cite{Sha53,EM79,Condon92,ZP96,CAHS03}.
It is therefore natural to use the same functions in the linear-time context of languages
and automata. 

Weighted automata with discounted sum have been considered in~\cite{DrosteR07},
with multiple discount factors and a boolean acceptance condition (Muller or B\"uchi); 
they are shown to be equivalent to a weighted monadic second-order logic with
discounting. Several other works have considered quantitative generalizations of 
languages, over finite words~\cite{DrosteGastin07}, over trees~\cite{DrosteKR08},
or using finite lattices~\cite{GurfinkelC03}, but none of these works
has addressed the expressiveness questions and closure properties for quantitative 
languages that are studied here.

The lattice automata of~\cite{KL07} map finite words to values from a finite
lattice. The lattice automata with B\"uchi condition are analogous to our 
$\LimSup$ automata, and their closure properties are established there.
However, the other classes of quantitative automata ($\Sum$, limit-average, 
discounted-sum) are not studied there as they cannot be defined using lattice 
operations and finite lattices.

\section{Quantitative Languages}\label{sec:definitions}

A \emph{quantitative language} $L$ over a finite alphabet $\Sigma$
is either a mapping $L: \Sigma^{+} \to \real$ or a mapping $L: \Sigma^{\omega} \to \real$, 
where $\real$ is the set of real numbers. 

\paragraph{\bf Weighted automata.}
A \emph{weighted automaton} is a tuple $A=\tuple{Q,q_I,\Sigma,\delta,\weight}$,
where
\begin{enumerate}[$\bullet$]
\item $Q$ is a finite set of states, $q_I \in Q$ is the initial state, and $\Sigma$ is a finite alphabet;
\item $\delta \subseteq Q \times \Sigma \times Q$ is a finite set of labelled transitions.
We assume that $\delta$ is \emph{total}, i.e., for all 
$q \in Q$ and $\sigma \in \Sigma$, there exists $q'$ such that 
$(q,\sigma,q') \in \delta$;
%%for at least one $q' \in Q$;
%\item $F \subseteq Q$ is the set of final (or accepting) states.
\item $\weight: \delta \to \rat$ is a \emph{weight} function, where $\rat$ is the 
set of rational numbers. We assume that rational numbers are encoded as pairs of 
integers in binary.
\end{enumerate}

\noindent We say that $A$ is \emph{deterministic} if for all 
$q \in Q$ and $\sigma \in \Sigma$, there exists $(q,\sigma,q') \in \delta$ 
for exactly one $q' \in Q$. We sometimes call automata \emph{nondeterministic}
to emphasize that they are not necessarily deterministic.
 
A \emph{run} of $A$ over a finite (resp. infinite) word $w=\sigma_1 \sigma_2 \dots$ 
is a finite (resp. infinite) sequence $r = q_0 \sigma_1 q_1 \sigma_2 \dots $ 
of states and letters such that 
\begin{compressEnum}
\itCompress $q_0 = q_I$, and
\itCompress $(q_i,\sigma_{i+1},q_{i+1}) \in \delta$ for all $0 \leq i < \abs{w}$.
%\itCompress $q_n \in F$.
\end{compressEnum}
We denote by $\weight(r) = v_0 v_1 \dots$ the sequence of weights that occur in~$r$ 
where $v_i = \weight(q_i,\sigma_{i+1},q_{i+1})$ for all $0 \leq i < \abs{w}$.

Given a \emph{value function} $\Val: \rat^+ \to \real$ (resp. $\Val: \rat^{\omega} \to \real$), 
we say that the $\Val$-automaton~$A$ defines the quantitative language $L_A$ such 
that for all $w \in \Sigma^{+}$ (resp. $w \in \Sigma^{\omega}$):
$$L_A(w) = \sup \{\Val(\weight(r)) \mid r \text{ is a run of } A \text{ over } w\}.$$
We assume that $\Val(v)$ is bounded when the numbers in $v$ are taken from a finite set (namely,
the set of weights in $A$), and since weighted automata are total, every word has
at least one run and thus $L_A(w)$ is not infinite.

%We consider the following functions.
%\paragraph{\bf Quantitative functions}  % Max, LimSup, LimInf, LimAvg, Disc$_{\lambda}$.
We consider the following value functions to define quantitative languages (they all satisfy 
the boundedness assumption above).
Given a finite sequence $v= v_1 \dots v_n$ of rational numbers, define
\begin{enumerate}[$\bullet$]
\item $\Maxf(v) = \max \{v_i \mid 1 \leq i \leq n\}$;\medskip
\item $\Last(v) = v_n$;\medskip
\item $\Sum(v) = \displaystyle\sum_{i=1}^{n} v_i$;
%\item $\Last(v) = v_n; \quad\quad \bullet\ \Sum(v) = \displaystyle\sum_{i=1}^{n} v_i$    
\end{enumerate}
% $$ \Last(v) = v_n, \quad\quad \Maxf(v) = \sup \{v_i \mid 1 \leq i \leq n\}, \quad\quad \Sum(v) = \displaystyle\sum_{i=1}^{n} v_i.$$
Given an infinite sequence $v=v_0 v_1 \dots$ of rational numbers, define
\begin{enumerate}[$\bullet$]
\item $\Max(v)    = \sup \{v_n \mid n \geq 0\}$;\medskip
\item $\LimSup(v) = \displaystyle\limsup_{n\to\infty} \ v_n = \lim_{n\to\infty} \sup \{v_i \mid i \geq n\}$;\medskip
\item $\LimInf(v) = \displaystyle\liminf_{n\to\infty} \ v_n = \lim_{n\to\infty} \inf \{v_i \mid i \geq n\}$;\medskip
\item $\LimAvg(v) = \displaystyle\liminf_{n\to\infty} \ \frac{1}{n} \cdot \sum_{i=0}^{n-1} v_i$;\medskip
\item for $0 < \lambda < 1$, $\Disc_{\lambda}(v) = \displaystyle \sum_{i=0}^{\infty} \lambda^i \cdot v_i$;
\end{enumerate}
Intuitively for a sequence $v=v_0 v_1 \dots$ of rational numbers from the finite set $V$,
the $\Max$ function chooses the maximal number that appear in $v$; 
the $\LimSup$ function chooses the maximal number that appear infinitely often in $v$;
the $\LimInf$ function chooses the minimal number that appear infinitely often in $v$;
the $\LimAvg$ functions gives the long-run average of the numbers in $v$; 
and the $\Disc_{\lambda}$ gives the discounted sum of the numbers in $v$.
%%For decision problems, we always assume that the discount
%%factor~$\lambda$ is a rational number. 
%
Note that $\LimAvg(v)$ is defined using $\liminf$ and is therefore well-defined;
all results of this paper hold also if the limit-average of $v$ is 
defined instead as 
$\limsup_{n\to\infty} \ \frac{1}{n}\cdot \sum_{i=0}^{n-1} v_i$.
One could also consider the value function $\inf \{v_n \mid n \geq 0\}$ 
and obtain results analogous to the $\Max$ value function.
Note that the classical finite-word acceptance condition of finite automata (defining regular languages)
can be encoded by $\Last$-automata with weights in $\{0,1\}$, while 
B\"uchi and coB\"uchi automata are special cases of respectively 
$\LimSup$- and $\LimInf$-automata, with weights in $\{0,1\}$. 
The class of languages defined by nondeterministic B\"uchi automata is called
$\omega$-regular.

\smallskip\noindent{\em Significance of value functions.}
%We now describe the significance of the value functions chosen. 
The value functions provide natural generalizations of the classical boolean 
languages, they are complete for different levels of the Borel hierarchy, and
they have been well studied in the context of game theory.
\begin{enumerate}[(1)]
\item The $\Max$ value function is the natural quantitative generalization of 
the reachability condition and is complete for the first level of the Borel 
hierarchy ($\Sigma_1$ complete).

\item The $\LimSup$ and $\LimInf$ objectives are the natural quantitative 
generalizations of the classical B\"uchi and coB\"uchi conditions. 
Moreover, the $\LimSup$ and $\LimInf$ objectives are complete for the second 
level of the Borel hierarchy, and hence important and canonical quantitative 
functions ($\LimSup$ and $\LimInf$ objectives are 
$\Pi_2$ and $\Sigma_2$ complete, respectively) (see~\cite{Wadge,MannaPnueliBook}
for details related to completeness and reducibility of objectives
in the Borel hierarchy).

\item The $\LimAvg$ and $\Disc_{\lambda}$ value functions have been studied in 
many different contexts in game theory.
Discounted functions on graph games were introduced in the seminal work of 
Shapley~\cite{Sha53}, and have been extensively studied in economics.
Discounted conditions have also been studied for discounting the future in 
systems theory~\cite{AHM03}. 
The $\LimAvg$ function has also been studied extensively
in the context of games on graphs: the works of Everett~\cite{Eve57}, 
Liggett-Lippman~\cite{LigLip69}, 
Hopfman-Karp~\cite{HofKar66}, Ehrenfeucht-Mycielski~\cite{EM79},  
Mertens-Neyman~\cite{MN81}, Zwick-Paterson~\cite{ZP96} have studied different 
classes of games with $\LimAvg$ objective.
Also see the books~\cite{FV97,Puterman} for applications of discounted 
and limit-average value functions in the context of games on graphs. 
Moreover, the $\LimAvg$ value function is complete for the third level of the 
Borel hierarchy ($\Pi_3$-complete)~\cite{ChaTCS07}.
\end{enumerate}
Hence the value functions considered are classical, canonical, and well-studied 
in the bran\-ching-time framework of games on graphs, and we study them in the 
linear-time framework of weighted automata.

% \begin{center}
% \begin{tabular}{lcl}
% $\Max(v) = \sup \{v_n \mid n \geq 0\}$; & & \\[+8pt]
% %\multicolumn{3}{c}{$\Max(v) = \sup \{v_n \mid n \geq 0\}$;}\\[+4pt]
% $\LimSup(v) = \displaystyle\limsup_{n\to\infty} \ v_n$; & $\quad\quad$ & $\LimInf(v) = \displaystyle\liminf_{n\to\infty} \ v_n$; \\
% $\LimAvg(v) = \displaystyle\liminf_{n\to\infty} \ \frac{1}{n} \sum_{i=0}^{n-1} v_i$; & $\quad\quad$ & For $0 < \lambda < 1$, $\Disc_{\lambda}(v) = \displaystyle \sum_{i=0}^{\infty} \lambda^i \cdot v_i$.
% \end{tabular}
% \end{center}

\paragraph{\bf Notation.} \label{page:notation}
Classes of weighted automata over infinite words are denoted with acronyms of the form 
$xy$ where $x$ is either {\sc N}(ondeterministic), {\sc D}(eterministic), 
or \nD\/ (when deterministic and nondeterministic automata have the same expressiveness),
% or {\sc A}(lter\-na\-ting) [see Section~\ref{sec:alternating}],
and $y$ is one of the following: {\sc Sup}, \Ls(LimSup), \Li(LimInf), \La(LimAvg), or \Di.
For B\"uchi and coB\"uchi condition, we use {\sc BW} and {\sc CW} respectively.
% (In {\sc Li} automata given a sequence $(v_i)_{i \geq 0}$ of rewards, the output is
% $\lim\inf_{i \to \infty} v_i$).
% A definition of B\"uchi and coB\"uchi automata can be found in~\cite{CDH08}.

\paragraph{\bf Reducibility.} 
A class $\C$ of weighted automata is \emph{reducible}
to a class $\C'$ of weighted automata if for every $A \in \C$ there exists 
$A' \in \C'$ such that $L_A=L_{A'}$, i.e., $L_{A}(w)=L_{A'}(w)$ for all (finite or infinite) words $w$.
In particular, a class of weighted automata \emph{can be determinized}
if it is reducible to its deterministic counterpart.
Reducibility relationships for (non)deterministic weighted automata
are given in~\cite{CDH08}.

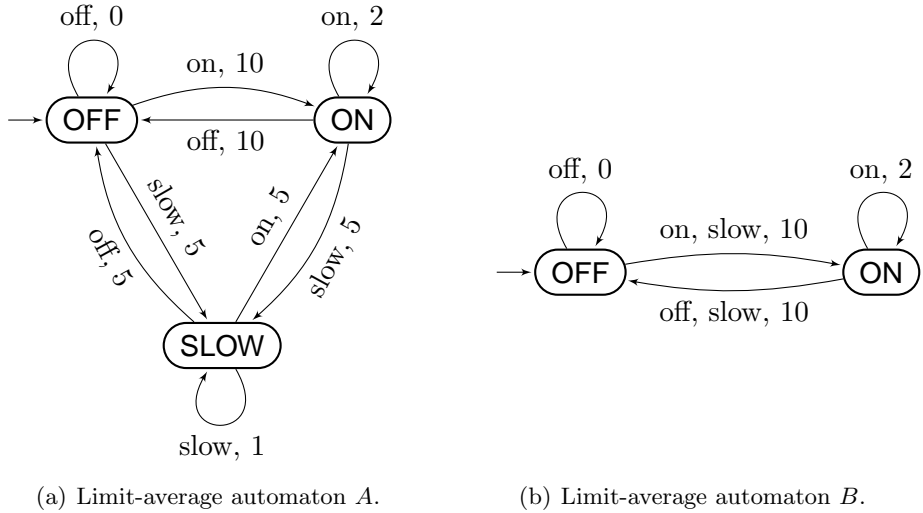
\begin{figure*}[!tb]
\begin{center}
     \subfigure[Limit-average automaton $A$. \label{fig:left}]{%
  \begin{tikzpicture}[node distance=1.8cm,auto,shorten >=1 pt,>=latex']
  \node[rdbox,initial,initial text={}]  (0)  at (150:2) {OFF} ;
  \node[rdbox]          (1)  at ( 30:2) {ON} ;
  \node[rdbox]          (2)  at (270:2) {SLOW} ;
  \draw[->] (0) edge[loop above,out=120, in=60,looseness=8] 
                node[above]       {off, 0}(0);
%                node[below]       {0}(0);
  \draw[->] (1) edge[loop above,out=120, in=60,looseness=8] 
                node[above]       {on, 2}(1);
%                node[below]       {2}(1);
  \draw[->] (2) edge[loop below,out=300, in=240,looseness=8] 
%                node[above]       {1}
                node[below]       {slow, 1}(2);
  \draw[->] (0) edge[bend left=20]  node{on, 10}  (1);
  \draw[->] (1) edge[bend left= 0]  node{off, 10} (0);
  \draw[->] (0) edge[bend left= 0]  node[sloped]{\mlap{slow, 5\quad}}  (2);
  \draw[->] (2) edge[bend left=20]  node[sloped]{\mlap{\quad off, 5}} (0);
  \draw[->] (1) edge[bend left=20]  node[sloped]{\mlap{slow, 5\quad}}  (2);
  \draw[->] (2) edge[bend left= 0]  node[sloped]{\mlap{\quad on, 5}} (1);
  \end{tikzpicture}\quad
}
     \subfigure[Limit-average automaton $B$. \label{fig:right}]{%
  \quad\begin{tikzpicture}[node distance=1.8cm,auto,shorten >=1 pt,>=latex']
  \node[rdbox,initial,initial text={}]  (0)  at (0,0) {OFF} ;
  \node[rdbox]          (1)  at (4,0) {ON} ;
  \node at(2,-2.5){};
  \draw[->] (0) edge[loop above,out=120, in=60,looseness=8] 
                node[above]       {off, 0}(0);
%                node[below]       {0}(0);
  \draw[->] (1) edge[loop above,out=120, in=60,looseness=8] 
                node[above]       {on, 2}(1);
%                node[below]       {2}(1);
  \draw[->] (0) edge[bend left=10]  node{on, slow, 10}  (1);
  \draw[->] (1) edge[bend left=10]  node{off, slow, 10} (0);
  \end{tikzpicture}
}
\end{center}
%\vspace{-5mm}
\caption{Specifications for the energy consumption of a motor:
$A$ refines $B$, i.e., $L_A \leq L_{B}$.\label{fig:motor-spec}}
\end{figure*}

\paragraph{\bf Composition.} 
Given two quantitative languages $L$ and $L'$ over $\Sigma$,
and a rational number $c$,
we denote by $\max(L,L')$ (resp. $\min(L,L')$, $L+L'$, $c+L$, and $cL$)
the quantitative language that assigns $\max\{L(w),L'(w)\}$
(resp. $\min\{L(w),L'(w)\}$, $L(w) + L'(w)$, $c+L(w)$, and $c\cdot L(w)$) to each word 
$w \in \Sigma^{+}$ (or $w \in \Sigma^{\omega}$). We say that $c+L$ is the \emph{shift by $c$}
of $L$ and that $cL$ is the \emph{scale by $c$} of $L$.
The language $1-L$ is called the \emph{complement} of $L$.
The $\max$, $\min$ and complement operators for quantitative languages 
generalize respectively the union, intersection and complement 
operator for boolean languages. For instance, De Morgan's laws hold
(the complement of the max of two languages is the min of their complement, etc.)
and complementing twice leave languages unchanged.

%\section{Examples}\label{sec-examples}
%We present three examples to illustrate how weighted automata are useful 
%to specify quantitative properties.

% \smallskip\noindent{\it Example 1.} Weighted automata  can be used to specify and verify reliability 
% requirements: if a special symbol $\bot$ is used to denote failure and has 
% weight~$1$, while the other symbols have weight $0$, then a limit-average 
% automaton can specify a bound on the rate of failure in the long run behavior
% of the system~\cite{CGHIKPS08}.

\smallskip\noindent{\it Example 1.} We consider a simple illustration of the use of limit-average automata
to model the energy consumption of a motor. The automaton $B$ in Figure~\ref{fig:right}
specifies the maximal energy consumption to maintain the motor on or off, and the 
maximal consumption for a mode change. The specification abstracts away that a mode
change can occur smoothly with the $\mathit{slow}$ command. A refined specification $A$ is 
given in Figure~\ref{fig:left} where the effect of slowing down is captured
by a third state. One can check that $L_{A}(w)\leq L_{B}(w)$ 
for all words~$w \in \{\mathit{on},\mathit{off},\mathit{slow}\}^\omega$.
Given two limit-average automata that model the energy consumption of 
two different motors, one needs to define composition operations for 
weighted automata to obtain the maximal, minimal, and sum of the average 
energy consumption of the motors.

\begin{figure*}[!t]
\begin{center}
     \subfigure[100 dollars invested in bank $A_1$. \label{fig:bank1}]{%
  \begin{tikzpicture}[node distance=1.8cm,auto,shorten >=1 pt,>=latex']
  \node[rdbox,initial,initial text={}]  (0)  at (0,0) {$G_1$} ;
  \node[rdbox]          (1)  at (3,0) {$B_1$} ;
  \draw[->] (0) edge[loop above,out=120, in=60,looseness=8] 
                node[above]       {$(g_1,\cdot),8$}(0);
%                node[below]       {0}(0);
  \draw[->] (1) edge[loop above,out=120, in=60,looseness=8] 
                node[above]       {$(b_1,\cdot),2$}(1);
%                node[below]       {2}(1);
  \draw[->] (0) edge[bend left=15]  node{$(b_1,\cdot),2$}  (1);
  \draw[->] (1) edge[bend left=15]  node{$(g_1,\cdot),8$} (0);
  \end{tikzpicture}\quad
}
     \subfigure[100 dollars invested in bank $A_2$. \label{fig:bank2}]{%
  \quad\begin{tikzpicture}[node distance=1.8cm,auto,shorten >=1 pt,>=latex']
  \node[rdbox,initial,initial text={}]  (0)  at (0,0) {$G_2$} ;
  \node[rdbox]          (1)  at (3,0) {$B_2$} ;
  \draw[->] (0) edge[loop above,out=120, in=60,looseness=8] 
                node[above]       {$(\cdot,g_2),6$}(0);
%                node[below]       {0}(0);
  \draw[->] (1) edge[loop above,out=120, in=60,looseness=8] 
                node[above]       {$(\cdot,b_2),4$}(1);
%                node[below]       {2}(1);
  \draw[->] (0) edge[bend left=15]  node{$(\cdot,b_2),4$}  (1);
  \draw[->] (1) edge[bend left=15]  node{$(\cdot,g_2),6$} (0);
  \end{tikzpicture}
}
\end{center}
%\vspace{-5mm}
\caption{The discounted-sum automaton models of two banks.\label{fig:bank-spec}}
\end{figure*}
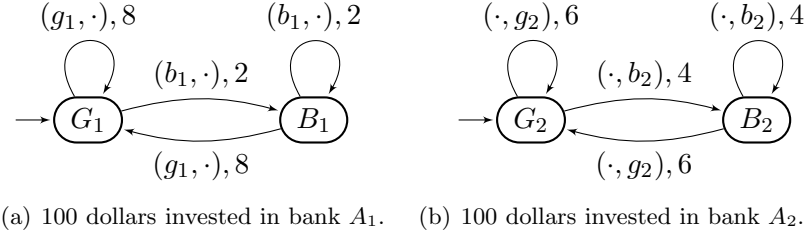

\smallskip\noindent{\it Example 2.} Consider an investment of 100~dollars
that can be made in two banks~$A_1$ and~$A_2$ as follows: 
(a)~100~dollars to bank~$A_1$, (b)~100~dollars to bank $A_2$, or
(c)~50~dollars to bank~$A_1$ and 50~dollars to bank~$A_2$.
The banks can be either in a good state (denoted $G_1$, $G_2$)
or in a bad state (denoted $B_1$, $B_2$).
If it is in a good state, then~$A_1$ offers 8\% reward while~$A_2$ offers 6\% reward.
If it is in a bad state, then~$A_1$ offers 2\% reward while~$A_2$ offers 4\% reward.
The change of state is triggered by the input symbols~$b_1, b_2$ (from a good
to a bad state) and~$g_1, g_2$ (from a bad to a good state).
The rewards received earlier weight more than rewards received later 
due to inflation represented by the discount factor.
The automata~$A_1$ and~$A_2$ in Figure~\ref{fig:bank-spec} specify the behavior 
of the two banks for an investment of 100~dollars, where the input alphabet is 
$\set{g_1,b_1} \times \set{g_2,b_2}$ (where the notation $(g_1,\cdot)$ 
represents the two letters $(g_1,g_2)$ and $(g_1,b_2)$, and similarly for the other symbols). 
If 50~dollars are invested in each bank, then we obtain automata~$C_1$ and~$C_2$ 
from~$A_1$ and~$A_2$ where each reward is halved. 
The combined automaton is obtained as the composition of~$C_1$ and~$C_2$ 
under the sum operator.

\section{Expressiveness Results} %%% for Weighted Automata}

The expressive power of weighted automata can be compared by mean of the
reducibility relation, saying that a class $\C$ of weighted automata  is at least
as expressive as a class $\C'$ if every quantitative language definable
by some automaton in $\C'$ is also definable by some automaton in $\C$.
The comparison includes boolean languages, considering them as a special case of quantitative languages
of the form $L:\Sigma^{\omega} \to \{0,1\}$.
It was shown in~\cite{CDH08} that a wide variety of classes of quantitative languages
can be defined by the different types of weighted automata, depending on the
value function and whether they are deterministic or not. This contrasts with
the situation for boolean languages where most of the classes of automata define
$\omega$-regular languages.
In this section, we investigate alternative ways of comparing the expressive power
of weighted automata and of classical finite automata. 
First, we use the cut-point languages of weighted automata to compare with
the class of $\omega$-regular languages, and then we use weighted automata
with boolean weights, i.e. all transitions have weight $0$ or $1$, to compare 
with general weighted automata.

\subsection{Cut-point languages}

Let $L$ be a quantitative language over infinite words and let $\eta \in \real$ be a threshold.
The \emph{cut-point language} defined by $(L,\eta)$ is the (boolean) language
$$L^{\geq \eta} = \{w \in \Sigma^{\omega} \mid L(w) \geq \eta\}.$$
Cut-point languages for finite words are defined analogously. They
have been first defined for probabilistic automata~\cite{Rabin63},
then generalized to inverse image recognition for semiring automata over finite words (see e.g.~\cite{KuichS86,CortesM00}).
It is easy to see that the cut-point languages of $\Maxf$- and $\Last$-automata are regular
(they have the same acceptance condition as finite automata),
those of $\Sum$-automata are context-free (using a stack to simulate accumulated weights), 
and those of $\Max$-, $\LimSup$-, and $\LimInf$-automata are $\omega$-regular
(they have the same acceptance condition as B\"uchi and coB\"uchi automata).

We show that the classes of cut-point languages definable by (non)deterministic limit-average and 
discounted-sum automata are incomparable with the $\omega$-regular languages. 
One direction of this result follows from Theorem~\ref{theo:cut-point-language}, 
and the other direction follows from~\cite[Theorems 13 and 14]{CDH08} where 
$\omega$-regular languages are given that are not definable as cut-point language 
of nondeterministic limit-average and discounted-sum automata.

\begin{thm}\label{theo:cut-point-language}
There exist deterministic limit-average and discounted-sum automata
whose cut-point language is not $\omega$-regular.
\end{thm}

\proof
Consider the alphabet $\Sigma=\set{a,b}$, and consider the languages $L_1$
that assigns to each word its long-run average number of $a$'s,
and $L_2$ that assigns the discounted sum of $a$'s. Note that $L_1$
is definable by a deterministic limit-average automaton,
and $L_2$ by a deterministic discounted-sum automaton.
It was shown in~\cite{Cha-TCS} that the cut-point language 
$L_1^{\geq 1}$ is complete for the third level of the Borel hierarchy,
and therefore is not $\omega$-regular.
We show that $L_2^{\geq 1}$ is not $\omega$-regular.

Given a finite word $w \in \Sigma^*$, let $\reward_a(w) = \sum_{i\mid w_i = a} \lambda^{i-1}$
be the discounted sum of $a$'s in $w$. We say that $w$ is \emph{ambiguous}
if $1- \frac{\lambda^{\abs{w}}}{1-\lambda} \leq \reward_a(w) < 1$. The ambiguity lies
in that some continuations of $w$ (namely $w.a^{\omega}$) are in $L_2^{\geq 1}$ and some are not
(namely $w.b^{\omega}$).
We show that for all $\lambda > \frac{1}{2}$, if $w$ is ambiguous,
then either $w.a$ or $w.b$ is ambiguous, which entails 
that there exists an infinite word $\hat{w}$ all of whose 
finite prefixes are ambiguous (and $L_2(\hat{w}) = 1$).
To do this, assume that $1- \frac{\lambda^{\abs{w}}}{1-\lambda} \leq \reward_a(w) < 1$,
and show that either
$1- \frac{\lambda^{1+\abs{w}}}{1-\lambda} \leq \reward_a(w.a) < 1$
or $1- \frac{\lambda^{1+\abs{w}}}{1-\lambda} \leq \reward_a(w.b) < 1$.
Since $\reward_a(w.a) = \reward_a(w) + \lambda^{\abs{w}}$ and $\reward_a(w.b) = \reward_a(w)$,
we have to show that 
$1- \frac{\lambda^{\abs{w}}}{1-\lambda} \leq \reward_a(w) < 1 - \lambda^{\abs{w}}$
or $1- \frac{\lambda^{1+\abs{w}}}{1-\lambda} \leq \reward_a(w) < 1$.
This holds if $1- \frac{\lambda^{1+\abs{w}}}{1-\lambda} < 1 - \lambda^{\abs{w}}$,
which is equivalent to $\lambda > \frac{1}{2}$.

Now, we show that if there exists a nondeterministic B\"uchi automaton~$A$ for~$L_2^{\geq 1}$,
then the set of states $S_n$ reached in~$A$ by reading the first~$n$ letters
of $\hat{w}$ (which we denote by $\hat{w}_{[1\dots n]}$) should be different for each~$n$, 
i.e., $n \neq m$ implies $S_n \neq S_m$.
Towards a contradiction, assume that $S_n = S_m$ for $n < m$.
Then for all continuations $w' \in \Sigma^\omega$, we have $\hat{w}_{[1\dots n]}.w' \in L_2^{\geq 1}$ 
if and only if $\hat{w}_{[1\dots m]}.w' \in L_2^{\geq 1}$ $(\star)$.

In particular, consider the continuations $\hat{w}_{[n+1\dots]}$ and $\hat{w}_{[m+1\dots]}$,
and for each $i \geq 1$, let $\gamma^i = \reward_a(\hat{w}_{[1\dots i]})$ and 
$K^i = L_2(\hat{w}_{[i+1\dots]})$. Then, we have $\gamma^i + \lambda^i \cdot K^i = 1$, 
and thus $\gamma^m + \lambda^m \cdot K^n \leq 1$ iff $K^n \leq K^m$. Since either $K^n \leq K^m$
or $K^m \leq K^n$, we have either $L_2(\hat{w}_{[1\dots m]}.\hat{w}_{[n+1\dots]}) \leq 1$ or 
$L_2(\hat{w}_{[1\dots n]}.\hat{w}_{[m+1\dots]}) \leq 1$. By $(\star)$,
this implies that either $L_2(\hat{w}_{[1\dots m]}.\hat{w}_{[n+1\dots]}) = 1$, or 
$L_2(\hat{w}_{[1\dots n]}.\hat{w}_{[m+1\dots]}) = 1$, and in both cases since 
$L_2(\hat{w}) = 1$, we get 
$$\frac{1-\gamma^m}{\lambda^m} = \frac{1-\gamma^n}{\lambda^n}.$$
This implies $\lambda^{m-n} (1-P(\lambda)) = 1 - Q(\lambda)$
where $P(\lambda) = \reward_a(\hat{w}_{[1\dots n]})$ and $Q(\lambda) = \reward_a(\hat{w}_{[1\dots m]})$
are polynomials of respective degree $n-1$ and $m-1$, and with coefficients in 
the set $\{0,1\}$. First, observe that the equation is not identically $0$
because the coefficient of the term of degree $0$ is not $0$ (as the first letter
of $\hat{w}$ must be $b$ since $a$ is not ambiguous). Second, every
coefficient in the equation is in the set $\{-2,-1,0,1,2\}$, and a classical
result shows that if $\frac{p}{q}$ is a solution of a polynomial equation
with $p$ and $q$ mutually prime, then $p$ divides the coefficient of degree $0$,
and $q$ divides the coefficient of highest degree. Therefore, no rational number 
in the interval $]\frac{1}{2},1[\,$ can be a solution.
This shows that $n \neq m$ implies $S_n \neq S_m$, and thus the automaton $A$ cannot have
finitely many states.
\qed

% Let $n \in \nat$, and let $w_1,w_2 \in \Sigma^n$.
% It is easy to see that if $1- \frac{\lambda^n}{1-\lambda} < \reward_a(w_1) < 1$ and $\reward_a(w_2) < \reward_a(w_1)$, then 
% there exists a continuation $w' \in \Sigma^\omega$ such that $L_2(w_1) \geq 1$
% and $L_2(w_2) < 1$. Since the set of values of words in $L_2$ is dense
% in the interval $[0,1]$ when $\lambda \geq \frac{1}{2}$, 

We note that cut-point languages are not stable under arbitrarily small 
perturbations of the transition weights, nor of the value of the cut-point. 
Consider the quantitative languages $L_1$, $L_2$ from the proof of Theorem~\ref{theo:cut-point-language}.
If for instance a limit-average automaton $A$ assigns weight $1+\epsilon$ 
to the $a$'s and $0$ to the $b$'s, its cut-point language $L_A^{\geq 1}$
is clearly different from $L_1^{\geq 1}$, no matter the value of $\epsilon > 0$. 
The same holds with respect to $L_2$ if $A$ is interpreted as a discounted-sum automaton.

In the theory of probabilistic automata, where finite words are assigned a probability
of acceptance, the cut-point languages may also be non-regular. Therefore, 
one considers the special case where the cut-point is isolated, and
shows that the cut-point languages are then regular~\cite{Rabin63}.

A number $\eta$ is an \emph{isolated cut-point} of a quantitative language $L$
if there exists $\epsilon > 0$ such that
$$\abs{L(w) - \eta} > \epsilon \text{ for all } w \in \Sigma^{\omega}.$$

% We show that the nondeterministic discounted-sum automata with isolated cut-point define
% an $\omega$-regular language, and the same holds for deterministic limit-average
% automata. 
We argue that isolated cut-point languages are robust, in that 
they remain unchanged under small perturbations of 
the transition weights. This follows from a more general result about
the robustness of weighted automata.

A class of weighted automata is robust if a small (syntactical) perturbation
in the weights of an automaton induces only a small (semantical) perturbation 
in the values of the words in the quantitative language of the automaton,
and the semantical perturbation tends to $0$ when the syntactical perturbation tends to $0$.
% Uniform robustness of a class of automata means that the semantical change in the 
% quantitative language depends only on the value of the syntactical change, 
% but not on the values of the weights, or on the transition relation 
% of each particular automata of the class. 
To formally define robustness, we need $\epsilon$-approximations of automata,
and distance between quantitative languages.

Let $A=\tuple{Q,q_I,\Sigma,\delta,\weight}$ be a (nondeterministic) weighted automaton,
and let $\epsilon \in \posreal$. We say that a weighted automaton $B=\tuple{Q',q'_I,\Sigma,\delta',\weight'}$
is an \emph{$\epsilon$-approximation} of $A$ if
\begin{enumerate}[$\bullet$]
\item $Q' = Q$, $q'_I = q_I$, $\delta' = \delta$, and
\item $\abs{\weight'(q,\sigma,q') - \weight(q,\sigma,q')} \leq \epsilon$ for all $(q,\sigma,q') \in \delta$.
\end{enumerate}
The \emph{$\sup$-distance} between two quantitative languages $L_1,L_2: \Sigma^{\omega} \to \real$
is defined by 
$$D_{\sup}(L_1,L_2) = \sup_{w \in \Sigma^{\omega}} \abs{L_1(w) - L_2(w)}.$$

We say that a class $\C$ of weighted automata is \emph{uniformly robust} if for all $\eta \in \sposreal$,
there exists $\epsilon \in \sposreal$ such that for all automata $A,B \in \C$ such that 
$B$ is an $\epsilon$-approximation of $A$, we have $D_{\sup}(L_A,L_B) \leq \eta$.
Note that uniform robustness implies a weaker notion of robustness where
a class $\C$ of weighted automata is called \emph{robust} if for all automata $A \in \C$
and for all $\eta \in \sposreal$, there exists $\epsilon \in \sposreal$ such that 
for all $\epsilon$-approximations $B$ of $A$ (with $B \in \C$), we have $D_{\sup}(L_A,L_B) \leq \eta$
(here the value of~$\epsilon$ can depend for instance on the weights of the automaton~$A$).

\begin{thm}\label{theo:uniformly-robust}
The classes of (non)deterministic $\Max$-, $\LimSup$-, $\LimInf$-, $\LimAvg$- and $\Disc$-automata
are uniformly robust.
\end{thm}

\proof
Let $A,B$ be two weighted automata with $B$ an $\epsilon$-approximation of $A$.
It is easy to see that for $\Max$-, $\LimSup$-, $\LimInf$- and $\LimAvg$-automata, 
the value of a run $r$ of $B$ differs by at most $\epsilon$ from the value of 
the same run in $A$. Therefore, $D_{\sup}(L_A,L_B) \leq \epsilon$ and we can take 
$\epsilon = \eta$. 
For $\Disc$-automata, the value of a run of $B$ differs by at most $\frac{\epsilon}{1-\lambda}$ 
from the value of the same run in $A$, where $\lambda$ is the discount factor. 
Therefore, we can take $\epsilon = \eta (1 - \lambda)$. 
\qed

As a corollary of Theorem~\ref{theo:uniformly-robust}, for an isolated cut-point $\eta$, 
the cut-point language $L^{\geq \eta}$ remains unchanged under small
perturbations of the weights. %%% transition weights. 

\begin{cor}
Let $L_A$ be the quantitative language defined by a weighted automaton $A$,
and let $\eta$ be an isolated cut-point of $L_A$. There exists a rational $\epsilon > 0$
such that for all $\epsilon$-approximations $B$ of $A$, we have $L_A^{\geq \eta} = L_B^{\geq \eta}$ 
(where $L_B$ is the quantitative language defined by $B$).
\end{cor}

Now, we show that the isolated cut-point languages of 
deterministic discounted-sum and limit-average automata are $\omega$-regular. 
For nondeterministic automata, the same property holds in the discounted-sum case, 
but the question is open for limit-average.

\begin{thm}
Let $L$ be the quantitative language defined by a $\Disc$-automaton.
If $\eta$ is an isolated cut-point of $L$, then the cut-point language $L^{\geq \eta}$
is $\omega$-regular.
%\footnote{It is even a "safety" language, i.e. definable by a DBW
%with all states accepting, except a sink state.}.
\end{thm}

\proof
Let~$\lambda$ be the discount factor of the $\Disc$-automaton~$A$ that defines~$L$.
Since~$\eta$ is an isolated cut-point of $L$, let $\epsilon > 0$ such that
$\abs{L(w) - \eta} > \epsilon$ for all $w \in \Sigma^{\omega}$.
Let $n \in \nat$ such that $u_n = \frac{V\cdot \lambda^n}{1-\lambda} < \epsilon $
where $V = \max_{(q,\sigma,q') \in \delta_A} \abs{\weight(q,\sigma,q')}$ is the largest weight in $A$.
Note that $u_n$ is a bound on the difference between the $\lambda$-discounted sum of the weights
in any infinite run $\hat{r}$ of $A$ and the $\lambda$-discounted sum of the weights in the prefix 
of length $n$ of $\hat{r}$, and $u_n \to 0$ when $n \to \infty$.

\noindent Consider an arbitrary run $r$ in $A$ of length $n$, and let $\weight(r)$ be the 
$\lambda$-discounted sum of the weights along $r$. Then, it should be clear that 
$\weight(r) \not\in [\eta - \epsilon + u_n, \eta + \epsilon - u_n]$, because
otherwise, the value of any (infinite) continuation of $r$ would lie in the
interval $[\eta - \epsilon, \eta + \epsilon]$, which would be a contradiction
to the fact that $\eta$ is an isolated cut-point of $L$.
Moreover, if $\weight(r) \leq \eta - \epsilon + u_n$, then any (infinite) continuation of $r$
has value less than $\eta - \epsilon + 2 u_n < \eta + \epsilon$, and therefore
less than $\eta$, while if $\weight(r) \geq  \eta + \epsilon - u_n$,
then any (infinite) continuation of $r$ has value greater than $\eta$.
Therefore, the cut-point language $L^{\geq \eta}$ can be defined by the
unfolding up to length $n$ of the $\Disc$-automaton that defines $L$, 
in which the states that are reached via a path with value at least
$\eta + \epsilon - u_n$ are declared to be accepting (for B\"uchi condition), 
and have a self-loop on $\Sigma$.
\qed

\begin{thm}
Let $L$ be the quantitative language defined by a deterministic $\LimAvg$-automaton.
If $\eta$ is an isolated cut-point of $L$, then the cut-point language $L^{\geq \eta}$
is $\omega$-regular.
%\footnote{It is even definable by a DBW.}.
\end{thm}

\proof
Let $A$ be a deterministic $\LimAvg$-automaton, defining the language $L$.
Consider the SCC-decomposition $C_1,C_2,\dots,C_k$ of the underlying graph of $A$.
For each $1 \leq i \leq k$, let $m_i$ and $M_i$ be the minimal and maximal
average weight of a cycle in $C_i$ (those values can be computed with
Karp's algorithm~\cite{Karp78}). 
It is easy to see that for every $1 \leq i \leq k$, for every $v \in [m_i, M_i]$, 
there exists a word $w \in \Sigma^{\omega}$ such that $L(w) = v$.
Therefore, since $\eta$ is an isolated cut-point of $L$, we have $\eta \not\in [m_i, M_i]$
for all $1 \leq i \leq k$. 
A deterministic B\"uchi automaton (\dbw) for $L^{\geq \eta}$ is obtained from $A$ by declaring to be accepting all 
states $q$ of $A$ such that $q \in C_i$ and $m_i > \eta$.
\qed

%\subsection{Robustness}

\subsection{Boolean weights}

We consider weighted automata with boolean set of weights, i.e. all transitions have weight $0$ or $1$.
The aim is to have a boolean counterpart to limit-average and discounted-sum
automata, and compare the expressive power.
We show that the restriction does not change the class of quantitative 
languages definable by limit-average automata, but does reduce the expressive 
power of discounted-sum automata.

%Given a set $R \subseteq \real$, and a class $\C$ of nondeterministic weighted automata,
%we denote by $\C_R$ the class of automata in $\C$ whose weights are rational numbers in~$R$.

\begin{thm}
The class of nondeterministic (resp., deterministic) $\LimAvg$-automata with rational weights in $[0,1]$
is reducible to the class of nondeterministic (resp., deterministic) $\LimAvg$-automata with weights $0$ and $1$ only.
% \nla${}_{[0,1]}$ is reducible to \nla${}_{\{0,1\}}$, and \dla${}_{[0,1]}$ is reducible to \dla${}_{\{0,1\}}$.
\end{thm}

\proof
Given a \nla\/ $A=\tuple{Q,q_I,\Sigma,\delta,\weight}$ with weights in $[0,1]$,
we construct a \nla\/ $B$ with weights in $\{0,1\}$ such that $L_A = L_B$. 

First, let $W = \{\weight(q,\sigma,q') \mid (q,\sigma,q') \in \delta\}$
be the set of weights that occur in $A$, and let $n_A$ be the smallest integer $n$
such that for all $v \in W$, there exists $p \in \nat$ such that $v = \frac{p}{n}$
(i.e., $\frac{1}{n_A}$ is the greatest common divisor of the weights of~$A$).
We define $B=\tuple{Q',q'_I,\Sigma,\delta',\weight'}$ as follows:
\begin{enumerate}[$\bullet$]
\item $Q' = Q \times [n_A]$ (where $[n_A]$ denotes the set $\{0,1,\dots,n_A-1\}$).
Intuitively, when we reach a state $(q,i)$ in $B$, it means that the state $q$
was reachable in $A$ and that the sum of the weights to reach $q$ is of the form
$k + \frac{i}{n_A}$ for some integer $k$. In $B$ however, the sum of the weights
to reach $(q,i)$ will then be $k$, and we store in the discrete state the information
that the remainder weight is $\frac{i}{n_A}$. Whenever this remainder exceeds $1$,
we introduce a weight $1$ and decrement the remainder.
\item $q'_I = (q_I,0)$;
\item for each transition $(q,\sigma,q') \in \delta$ and each value $i \in [n_A]$,
the following transitions are in $\delta'$ (where $v = \weight(q,\sigma,q')$):
	\begin{enumerate}[$-$]
	\item $((q,i),\sigma,(q',j))$ for $j=i+v\cdot n_A$ if $\frac{i}{n_A} + v < 1$; 
		the weight of such a transition is $0$ in $\weight'$,
	\item $((q,i),\sigma,(q',j))$ for $j=i+(v-1)\cdot n_A$ if $\frac{i}{n_A} + v \geq 1$; 
		the weight of such a transition is $1$ in $\weight'$.
	\end{enumerate}
Note that in the above, $v\cdot n_A$ is an integer and $j \in [n_A]$.
\end{enumerate}

There is a straightforward correspondence between the runs in $A$ and the runs in $B$.
Moreover, if the average weight of a prefix of length $n$ of a run in $A$ is $\frac{S}{n}$,
then the average weight of the prefix of length $n$ of the corresponding run in $B$
is between $\frac{S}{n}$ and $\frac{S+1}{n}$. Hence the difference tends to $0$
when $n \to \infty$. Therefore, the value of a run in $A$ is the same as the
value of the corresponding run in $B$, and therefore $L_A = L_B$.

Finally, note that if $A$ is deterministic, then $B$ is deterministic.
\qed

\begin{thm}
The class of deterministic $\Disc$-automata with rational weights in $[0,1]$
is not reducible to the class of (even nondeterministic) $\Disc$-automata with weights $0$ and $1$ only.
% ,for all rational discount factors different from $\frac{1}{2}$ and $\frac{2}{3}$.
\end{thm}

%\begin{thm}
%\nDi${}_{[0,1]}$ is not reducible to \nDi${}_{\{0,1\}}$ for all rational
%discount factors different from $\frac{1}{2}$ and $\frac{2}{3}$.
%\end{thm}

%{\bf New proof}

\proof
Given a discount factor $0 < \lambda <1$, consider the \ddi\/ over $\Sigma = \{a,b\}$ 
that consists of a single state with a self-loop over $a$ with weight $\frac{1+\lambda}{2}$ 
and a self-loop over $b$ with weight $0$. 
Let $L_{\lambda}$ be the quantitative language defined by this automaton.
Towards a contradiction, assume that this language is defined by a \nDi\/ $A$ with weights in $\{0,1\}$.
First, consider the word $a b^{\omega}$ whose value in $L_{\lambda}$ is $\frac{1+\lambda}{2} < 1$. 
This entails that $A$ cannot have a transition from the initial state over $a$ with weight $1$ 
(as this would imply that $L_A(a b^{\omega}) \geq 1$).
Now, the maximal value that $L_A$ can assign to the word $a^{\omega}$ is 
$\lambda + \lambda^2 + \lambda^3 + \cdots = \frac{\lambda}{1-\lambda}$ which is
strictly smaller than $L_{\lambda}(a^{\omega}) = \frac{1+\lambda}{2(1-\lambda)}$.
This shows that $A$ cannot exist.
\qed

\section{Closure Properties} %%% of Weighted Automata}
\label{sec:closure-properties}

We study the closure properties of weighted automata
with respect to $\max$, $\min$, complement and sum. We say that a class $\C$ of weighted
automata is \emph{closed} under a binary operator $\op(\cdot,\cdot)$ 
(resp. a unary operator $\op'(\cdot)$) if for all $A_1,A_2 \in \C$,
there exists $A_{12} \in \C$ such that $L_{A_{12}} = \op(L_{A_1},L_{A_2})$
(resp. $L_{A_{12}} = \op'(L_{A_1})$).
All closure properties that we present in this paper are constructive:
when $\C$ is closed under an operator,
we can always construct the automaton $A_{12} \in \C$ given $A_1,A_2 \in \C$.
We say that the \emph{cost} of the closure property of $\C$ under a binary operator $\op$ 
is at most $O(f(n_1,m_1,n_2,m_2))$ if 
for all automata $A_1,A_2 \in \C$ with $n_i$ states and $m_i$ transitions (for $i=1,2$ respectively),
the constructed automaton $A_{12} \in \C$ such that $L_{A_{12}} = \op(L_{A_1},L_{A_2})$
has at most $O(f(n_1,m_1,n_2,m_2))$ many states.
Analogously, the \emph{cost} of the closure property of $\C$ under a unary operator $\op'$ 
is at most $O(f(n,m))$ if 
for all automata $A_1 \in \C$ with $n$ states and $m$ transitions,
the constructed automaton $A_{12} \in \C$ such that $L_{A_{12}} = \op'(L_{A_1})$
has at most $O(f(n,m))$ many states.
For all reductions presented, the size of the largest weight in $A_{12}$ 
is linear in the size~$p$ of the largest weight in~$A_1,A_2$ (however, 
the time needed to compute the weights is quadratic in $p$,
as we need addition, multiplication, or comparison, which are quadratic
in $p$).

Notice that every class of weighted automata is closed under shift by~$c$ and
under scale by~$\abs{c}$ for all~$c \in \rat$. For $\Sum$-automata
and discounted-sum automata, we can define the shift by~$c$ by making a copy of the initial
states and adding $c$ to the weights of all its outgoing transitions.
For the other automata, it suffices to add~$c$
to (resp. multiply by~$\abs{c}$) all weights of an automaton to obtain the
automaton for the shift by~$c$ (resp. scale by~$\abs{c}$) of its language. 
Therefore, all closure properties also hold if the complement of a 
quantitative language~$L$ was defined as~$k-L$ for any constant~$k$.

Our purpose is the study of quantitative languages over infinite words.
For the sake of completeness we first give an overview of the closure 
properties for finite words. Table~\ref{tab:closure-properties}(a)
summarizes the closure properties for finite words,
and Table~\ref{tab:closure-properties}(b) for infinite words.

\subsection{Closure properties for finite words}

For finite words, we consider closure under $\max$, $\min$, complement,
and sum for $\Maxf$-, $\Last$- and $\Sum$-automata.

\begin{thm}\label{theo:max-closure-finite}
Deterministic $\Maxf$- and $\Last$-automata are closed under $\max$,
with cost $O(n_1 \cdot n_2)$.
Nondeterministic $\Maxf$-, $\Last$- and $\Sum$-automata are closed under $\max$,
with cost $O(n_1 + n_2)$.
Deterministic $\Sum$-automata are not closed under $\max$.
\end{thm}

\proof
For the nondeterministic automata, the result follows from the fact that 
the $\max$ operator can be obtained by an initial nondeterministic choice 
between two quantitative automata. For deterministic $\Maxf$- and $\Last$-automata,
the result is obtained using a standard synchronized product construction, 
where the weight of a transition in the product is the maximum of the corresponding
transition weights in the two automata.
Finally, deterministic $\Sum$-automata are not closed under the $\max$ operator 
because the language over $\Sigma = \{a,b\}$ that assigns to each finite word $w \in \Sigma^{+}$ 
the number $\max\{L_a(w),L_b(w)\}$ where $L_{\sigma}(w)$ is the number of occurrences of $\sigma$ in $w$ (for $\sigma = a,b$)
is definable by the max of two deterministic-$\Sum$ languages, but not 
by a deterministic $\Sum$-automaton (Theorem~2 in~\cite{CDH08}).
\qed

\begin{thm}\label{theo:min-closure-finite}
Deterministic and nondeterministic $\Maxf$-automata
are closed under $\min$, with cost $O(n_1\cdot m_1 \cdot n_2 \cdot m_2)$.
Deterministic and nondeterministic $\Last$-automata
are closed under $\min$, with cost $O(n_1 \cdot n_2)$.
Deterministic and nondeterministic $\Sum$-automata are not closed
under $\min$.
\end{thm}

\proof
Given two $\Last$-automata $A_1$ and $A_2$ (over the same alphabet), 
we use the classical synchronized product $A_{12} = A_1 \times A_2$, 
where the weight of a transition in $A_{12}$ is the minimum of the corresponding
transition weights in $A_1$ and $A_2$. It is easy to see that $L_{A_{12}} = \min(L_{A_1}, L_{A_2})$.
If $A_1$ and $A_2$ are deterministic, then so is $A_{12}$.

The construction for $\Maxf$-automata is the same as for $\Max$-automata
over infinite words given in the proof of Theorem~\ref{theo:max-closed-under-min}.

Finally, for $\Sum$-automata, consider the language $L_m$ over $\Sigma = \{a,b\}$ 
that assigns to each finite word $w \in \Sigma^{+}$ the value $\min\{L_a(w),L_b(w)\}$ 
where $L_{\sigma}(w)$ is the number of occurrences of $\sigma$ in $w$ (for $\sigma = a,b$).
We claim that $L_m$ is not definable by a nondeterministic $\Sum$-automaton.
Indeed, assume that the $\Sum$-automaton $A$ with state space $Q$ defines $L_m$.
First, the sum of weights in every reachable cycle of $A$ over $a$'s 
must be at most $0$. Otherwise, we can reach the cycle with a finite word $w_1$
and obtain an arbitrarily large value for the word $w_1 a^i$ for $i$ sufficiently 
large, while for any such $i$ the value of $w_1 a^i$ is the number of $b$'s
in $w_1$ which is independent of $i$. Analogously, the sum of weights in 
every reachable cycle of $A$ over $b$'s must be at most $0$.
Now, let $\beta= \max_{e \in \delta} \abs{\weight(e)}$ be the maximal 
weight in $A$, and consider the word $w = a^n b^n$ for $n > 2\beta \cdot \abs{Q}$.
Every run of $A$ over $a^n$ (or over $b^n$) can be decomposed in possibly nested cycles 
(since $A$ is nondeterministic) and a remaining non-cyclic path of length
at most $\abs{Q}$. Hence, the value of any run over $w$ is at most $2\beta \cdot \abs{Q}$.
However, the value of $w$ should be $n$, thus $A$ cannot exist.
\qed

\begin{thm}\label{theo:closure-under-complement-finite}
Deterministic $\Last$- and $\Sum$-automata are closed under complement, 
with cost $O(n)$.
Nondeterministic $\Last$-automata are closed under complement, 
with cost $O(2^n)$.
Nondeterministic $\Sum$ automata, and both deterministic and nondeterministic 
$\Maxf$-auto\-mata are not closed under complement.
\end{thm}

\proof
To define the complement of the language of a deterministic $\Sum$- (or $\Last$-) automaton, 
it suffices to multiply all the weights by $-1$, and then shift the language by $1$.
For the class of nondeterministic $\Last$-automata, the result follows from the fact that 
it is reducible to its deterministic counterpart.

The negative result for $\Maxf$-automata follows from an analogous in the boolean
case (consider the language $L$ over $\{a,b\}$ such that $L(a^i) = 0$
for all $i \geq 1$, and $L(w) = 1$ for all words containing the letter $b$). 
Finally, according to the proof of Theorem~\ref{theo:min-closure-finite}, 
the language $\min(L_a,L_b)$ where $L_{\sigma}(w)$ is the number of occurrences 
of $\sigma$ in $w$ (for $\sigma = a,b$) is not definable by a nondeterministic 
$\Sum$-automaton. Since $\min(L_a,L_b) = 1-\max(1-L_a,1-L_b)$ and 
\begin{compressEnum}
\itCompress $1-L_a$ and $1-L_b$ are definable by $\Sum$-automata, and
\itCompress nondeterministic $\Sum$-automata are closed under $\max$ (Theorem~\ref{theo:max-closure-finite}),
\end{compressEnum}
the language $\max(1-L_a,1-L_b)$ is definable by a nondeterministic $\Sum$-automaton,
but not its complement and the result follows.
\qed

\begin{thm}\label{theo:closure-under-sum-finite}
Every class of weighted automata over finite words is closed under sum.
The cost is $O(n_1\cdot n_2)$ for $\Last$- and $\Sum$-automata,
and $O(n_1\cdot m_1 \cdot n_2 \cdot m_2)$ for $\Maxf$-automata.
\end{thm}

\proof
It is easy to see that the synchronized product of two $\Last$-automata (resp. $\Sum$-automata)
defines the sum of their languages if the weight of a joint transition
is defined as the sum of the weights of the corresponding transitions in the two 
$\Last$-automata (resp. $\Sum$-automata).

We give the construction for two $\Maxf$-automata
% is the same as for $\Max$-automata over infinite words given in the proof of Theorem~\ref{theo:max-closed-under-sum}.
$A_1=\tuple{Q_1,q_I^1,\Sigma,\delta_1,\weight_1}$ and $A_2=\tuple{Q_2,q_I^2,\Sigma,\delta_2,\weight_2}$. 
We construct a $\Maxf$-automaton $A_{12}=\tuple{Q,q_I,\Sigma,\delta,\weight}$ such that
$L_{A_{12}} = L_{A_1} + L_{A_2}$. 
Let $V_i = \{\weight_i(e) \mid e \in \delta_i\}$ be the set of weights that appear in $A_i$ (for $i=1,2$),
%Let $V_1 \cup V_2 = \{v_1,\dots,v_n\}$
and define:
\begin{enumerate}[$\bullet$]
\item $Q = Q_1 \times V_1 \times Q_2 \times V_2$. Intuitively, we remember in 
a state $(q_1, v_1, q_2, v_2)$ the largest weights $v_1,v_2$ seen so far in the corresponding
runs of $A_1$ and $A_2$;
\item $q_I = (q_I^1, v_{\min}^1, q_I^2, v_{\min}^2)$ where $v_{\min}^i$ is the minimal weight in $V_i$ (for $i=1,2$);
\item For each $\sigma \in \Sigma$, the set $\delta$ contains all the triples 
$\tuple{(q_1, v_1, q_2, v_2),\sigma,(q'_1,v'_1,q'_2,v'_2)}$ such that 
$v_i \in V_i$, $(q_i,\sigma,q'_i) \in \delta_i$, and $v'_i = \max\{v_i,\weight(q_i,\sigma,q'_i)\}$,
for $i=1,2$;
\item $\weight$ is defined by \[\weight(\tuple{(q_1, v_1, q_2, v_2),\sigma,(q'_1,v'_1,q'_2,v'_2)}) = v'_1 + v'_2\]
for each $\tuple{(q_1, v_1, q_2, v_2),\sigma,(q'_1,v'_1,q'_2,v'_2)} \in \delta$.
\end{enumerate}
If $A_1$ and $A_2$ are deterministic, then $A_{12}$ is deterministic.
The result for deterministic $\Maxf$-automata follows.
\qed

% % % % % % % % % % % % % % % % % % % % % % % % % % % % % % % % % % % % % % % % % % % % % % % % % % % % % % % % % % % % 
\subsection{Closure under $\max$ for infinite words}
% % % % % % % % % % % % % % % % % % % % % % % % % % % % % % % % % % % % % % % % % % % % % % % % % % % % % % % % % % % % 

%The $\max$ operator generalizes the union of boolean languages.
The maximum of two quantitative languages defined by nondeterministic automata
can be obtained by an initial nondeterministic choice between the two automata.
This observation was also made in~\cite{DrosteR07} for discounted-sum automata.
For deterministic automata, a synchronized product can be used
for $\Max$ and $\LimSup$, while for $\LimInf$ we use the fact that \nli\/ 
is determinizable with an exponential blow-up~\cite{CDH08}.

% \begin{thm}\label{theo:max-closure}
% \nmax, \nli, \nls, \nla\/ and \nDi\/ are closed under $\max$, with cost $O(n_1+n_2)$.
% \dmax\/ and \dls\/ are closed under $\max$, with cost $O(n_1 \cdot n_2)$.
% \dli\/ is closed under $\max$ with cost $O((m_1+m_2)\cdot 2^{n_1 + n_2})$.
% \end{thm}

\begin{thm}\label{theo:max-closure}
The nondeterministic $\Max$-, $\LimSup$-, $\LimInf$-, $\LimAvg$- and $\Disc$-automata
are closed under $\max$, with cost $O(n_1+n_2)$,
the deterministic $\Max$- and $\LimSup$-automata with cost $O(n_1 \cdot n_2)$,
the deterministic $\LimInf$-automata with cost $O((m_1+m_2)^{n_1 + n_2})$.
\end{thm}

% \proof[Sketch]
% For all the nondeterministic quantitative automata, the result 
% follows from the fact that the $\max$ operator can be achieved
% with an initial nondeterministic choice between two quantitative 
% automata.
% Since \nli\/ is reducible to \dli\/ and \nmax\/ is reducible to \dmax\/~\cite{CDH08}, 
% the result follows for these classes.
% We now prove that \dls\/ is closed under the max operator.
% Given two \dls's $A_1$ and $A_2$ over the same alphabet, we construct the usual
% synchronized product $A_{12} = A_1 \times A_2$, where the weight of a
% transition in $A_{12}$ is the maximum of the corresponding
% transition weights in $A_1$ and $A_2$.
% It is easy to see that $L_{A_{12}} = \max(L_{A_1}, L_{A_2})$.
% \qed
\proof[Sketch]
For all the nondeterministic quantitative automata, the result 
follows from the fact that the $\max$ operator can be achieved
with an initial nondeterministic choice between two weighted 
automata.
For \dli, the result follows from the reducibility of \nli\/ to \dli\/
with an exponential blow-up~\cite{CDH08}.
We now prove that \dls\/ and \dmax\/ are closed under $\max$ with cost $O(n_1 \cdot n_2)$.
Given two \dls\/ (or \dmax) $A_1$ and $A_2$ over the same alphabet, we construct the usual
synchronized product $A_{12} = A_1 \times A_2$, where the weight of a
transition in $A_{12}$ is the maximum of the corresponding
transition weights in $A_1$ and $A_2$.
It is easy to see that $L_{A_{12}} = \max(L_{A_1}, L_{A_2})$ in both cases.
\qed

% \begin{thm}
% \dla and \ddi\/ are not closed under $\max$.
% \end{thm}

\begin{thm}
The deterministic $\LimAvg$- and $\Disc$-automata are not closed under $\max$.
\end{thm}

\proof 
The fact that \ddi\/ is not closed under $\max$ follows
from the proof of Theorem~16 in~\cite{CDH08}, where it is shown that the quantitative language
$\max(L_1,L_2)$ cannot be defined by a \ddi, where $L_1$ (resp. $L_2$) is the language
defined by the \ddi\/ that assigns weight $1$ (resp. $0$) to $a$'s 
and weight $0$ (resp. $1$) to $b$'s. 

We now show that \dla\/ is not closed under $\max$.
Consider the alphabet $\Sigma=\set{a,b}$ and the quantitative languages
$L_a$ and $L_b$ that assign the value of long-run average
number of $a$'s and $b$'s, respectively.
There exists \dla\/ for $L_a$ and $L_b$.
We show that $L_m=\max(L_a,L_b)$ cannot be expressed by 
a \dla. By contradiction, assume that $A$ is a \dla\/
with set of states $Q$ that defines $L_m$.
Consider any reachable cycle $C$ over $a$'s in $A$. The sum of the weights 
of the cycle must be its length $\abs{C}$, as if we consider the 
word $w^*=w_C \cdot (a^{\abs{C}})^\omega$ where $w_C$
is a finite word whose run reaches $C$, the value
of $w^*$ in $L_m$ is $1$. It follows that the sum of the weights
of the cycle $C$ must be $\abs{C}$. Hence, the sum of the weights
of all the reachable cycles $C$ over $a$'s in $A$ is $\abs{C}$.

\newcommand{\wh}{\widehat}

Consider the infinite word $w_\infty=(a^{\abs{Q}} \cdot b^{2\abs{Q}})^\omega$,
and let $w_j=(a^{\abs{Q}} \cdot b^{2\abs{Q}})^j$.
Since $L_m(w_\infty)=\frac{2}{3}$, the run of $A$ over $w_\infty$ 
has value $\frac{2}{3}$.
It follows that for all $\varepsilon>0$, there is an integer $j_\varepsilon$, such that
for all $j\geq j_\varepsilon$, we have 
\[
\frac{\weight(w_j)}{\abs{w_j}} \geq \frac{2}{3} -\varepsilon
\]
where $\weight(w_j)$ is the sum of the weights of the run of $A$ over $w_j$.
Consider a word $\wh{w}_\infty$ constructed as follows. We start 
with the empty word $\wh{w}_0$ and the initial state $q_0$ of $A$, and for all $j\geq 0$, we construct 
$(\wh{w}_{j+1}, q_{j+1})$ from $(\wh{w}_j,q_j)$ as follows: the state $q_{j+1}$
is the last state of the run of $A$ from $q_j$ over $a^{\abs{Q}} \cdot b^{2\abs{Q}}$. 
This run has to contain a cycle $C_{j+1}$ over $a$'s. We set 
$\wh{w}_{j+1} = \wh{w}_j \cdot a^{\abs{Q} + \abs{C_{j+1}}} \cdot b^{2\abs{Q}}$.
Observe that for all $j \geq 1$, the run of $A$ over $w_\infty$
in the segment between $w_j$ and $w_{j+1}$ is identical to
the run from $q_j$ to $q_{j+1}$ up to the repetition of the cycle $C_{j+1}$
once more.
The word $\wh{w}_\infty$ is the limit of this construction ($\wh{w}_j$ is
a prefix of $\wh{w}_\infty$ for all $j \geq 0$).
Let $\alpha_j=\sum_{i=1}^j \abs{C_i}$.    %%, where $C_i$ is the cycle in the $i$-th round.
Since $1\leq \abs{C_i} \leq \abs{Q}$ we have $j \leq \alpha_j \leq j \cdot \abs{Q}$.
Hence we have the following equality:
$\frac{\weight(\wh{w}_j)}{\abs{\wh{w}_j}} 
=\frac{\weight(w_j) + \alpha_j}{\abs{w_j}+ \alpha_j}$.
Hence for all $\varepsilon>0$, there exists $j_\varepsilon$ such that 
for all $j \geq j_\varepsilon$ we have
\[
\begin{array}{rcl}
\displaystyle
\frac{\weight(\wh{w}_j)}{\abs{\wh{w}_j}} 
& \geq  & 
\displaystyle 
\frac{ \frac{2}{3}\cdot \abs{w_j}   - \varepsilon \cdot \abs{w_j} + \alpha_j }{ \abs
{w_j} + \alpha_j} \\[2ex]
& \geq & 
\displaystyle 
\frac{2}{3} -\varepsilon + \frac{1}{3} \cdot \frac{\alpha_j}{\abs{w_j} + \alpha_j} 
\\[1ex]
& \geq & 
\displaystyle 
\frac{2}{3} -\varepsilon + \frac{1}{3} \cdot \frac{j}{j \cdot(3 \abs{Q} + \abs
{Q})} \\[1ex]
& \geq & 
\displaystyle 
\frac{2}{3} -\varepsilon + \frac{1}{12 \abs{Q}}. \\[1ex]
\end{array}
\] 

This shows that $\liminf_{j \to \infty} \frac{\weight(\wh{w}_j)}{\abs{\wh{w}_j}} \geq \frac{2}{3} + \frac{1}{12 \abs{Q}}$
and thus we have $L_A(\wh{w}_\infty) \geq \frac{2}{3} + \frac{1}{12 \abs{Q}}$.
Since $1 \leq \abs{C_i} \leq \abs{Q}$ for all $i\geq 1$, we have $L_m(\wh{w}_\infty)\leq\frac{2}{3}$
which is a contradiction.
\qed

% % % % % % % % % % % % % % % % % % % % % % % % % % % % % % % % % % % % % % % % % % % % % % % % % % % % % % % % % % % % 
\subsection{Closure under $\min$ for infinite words}
% % % % % % % % % % % % % % % % % % % % % % % % % % % % % % % % % % % % % % % % % % % % % % % % % % % % % % % % % % % % 

The positive results about closure properties under $\min$ for quantitative languages
generalize the closure properties of boolean languages under intersection.
The constructions are straightforward extensions of the standard constructions 
for finite, B\"uchi, and coB\"uchi automata (see e.g.~\cite{Vardi96}).

\begin{thm}\label{theo:max-closed-under-min}
% \dmax\/ and \nmax\/ are closed under the min operator, with cost $O(n_1\cdot m_1 \cdot n_2 \cdot m_2)$.
The (non)deterministic $\Max$-automata are closed under $\min$, with cost $O(n_1\cdot m_1 \cdot n_2 \cdot m_2)$,
\end{thm}

\proof
The construction in the proof of Theorem~\ref{theo:closure-under-sum-finite}
can be adapted by defining the weight $\weight(\tuple{(q_1, v_1, q_2, v_2),\sigma,(q'_1,v'_1,q'_2,v'_2)})$
as $\min\{v'_1, v'_2\}$ for each $\tuple{(q_1, v_1, q_2, v_2),\sigma,(q'_1,v'_1,q'_2,v'_2)} \in \delta$.
\qed

\begin{thm}
% \dls\/ is closed under $\min$, with cost $O(n_1 \cdot n_2 \cdot 2^{m_1 + m_2})$.
The deterministic $\LimSup$-automata are closed under $\min$ with cost $O(n_1 \cdot n_2 \cdot 2^{m_1 + m_2})$.
\end{thm}

\proof
Let $A_1=\tuple{Q_1,q_I^1,\Sigma,\delta_1,\weight_1}$ and $A_2=\tuple{Q_2,q_I^2,\Sigma,\delta_2,\weight_2}$
be two \dls. We construct a \dls\/ $A=\tuple{Q,q_I,\Sigma,\delta,\weight}$ such that
$L_A = \min\{L_{A_1},L_{A_2}\}$. Let $V_i = \{\weight_i(e) \mid e \in \delta_i\}$
be the set of weights that occur in $A_i$ (for $i=1,2$). For each weight
$v \in V_1 \cup V_2 = \{v_1,\dots,v_n\}$, we construct a \dbw\/ $A^v_{12}$ that 
consists of a copy of $A_1$ and a copy of $A_2$.
We switch from one copy to the other whenever an edge with weight at least $v$ 
is crossed. All such switching edges are accepting in $A^v_{12}$ (i.e., they have
weight~$1$ while all other edges have weight~$0$). The automaton $A$ then
consists of the synchronized product of these \dbw, where the weight of a joint
edge is the largest weight $v$ for which the underlying edge in $A^v_{12}$ is accepting.
Formally, let
\begin{enumerate}[$\bullet$]
\item $Q = Q_1 \times Q_2 \times \{1,2\}^m$ where $m = \abs{V_1 \cup V_2}$ (and assume $V_1 \cup V_2= \{v_1,\dots,v_m\}$);
\item $q_I = (q_I^1, q_I^2, b_1, \dots, b_m)$ where $b_i=1$ for all $1 \leq i \leq m$;
\item $\delta$ contains all the triples 
$(\tuple{q_1,q_2,b_1,\dots,b_m},\sigma, \tuple{q'_1,q'_2,b'_1,\dots,b'_m})$ such that
$\sigma \in \Sigma$ and 
	\begin{enumerate}[$-$]
	\item $(q_i,\sigma,q'_i) \in \delta_i$ for $i=1,2$;
	\item for all $1\leq j \leq m$, we have $b'_{j} = 3-b_j$ if 
$\weight_{b_j}(q_{b_j},\sigma,q'_{b_j}) \geq v_j$, and $b'_j = b_j$ otherwise.
	\end{enumerate}
\item $\weight$ assigns to each transition 
$(\tuple{q_1,q_2,b_1,\dots,b_m},\sigma, \tuple{q'_1,q'_2,b'_1,\dots,b'_m}) \in \delta$ 
the weight $v=\max(\{v_{\min}\}\cup \{v_j \mid b_j\neq b'_j\})$
where $v_{\min}$ is the minimal weight in $V_1 \cup V_2$.\qed
\end{enumerate}

\begin{thm}
% \dli\/ and \nli\/ are closed under $\min$, with cost $O(n_1 \cdot n_2)$.
% \nls\/ is closed under $\min$, with cost $O(n_1\cdot n_2 \cdot (m_1 + m_2))$.
The (non)deterministic $\LimInf$-automata are closed under $\min$ with cost $O(n_1 \cdot n_2)$, and
the nondeterministic $\LimSup$-automata with cost $O(n_1\cdot n_2 \cdot (m_1 + m_2))$.
\end{thm}

\proof
Let $A_1=\tuple{Q_1,q_I^1,\Sigma,\delta_1,\weight_1}$ and $A_2=\tuple{Q_2,q_I^2,\Sigma,\delta_2,\weight_2}$
be two \nls. We construct a \nls\/ $A=\tuple{Q,q_I,\Sigma,\delta,\weight}$ such that
$L_A = \min\{L_{A_1},L_{A_2}\}$. Let $V_i = \{\weight_i(e) \mid e \in \delta_i\}$
be the set of weights that appear in $A_i$ (for $i=1,2$). Let $V_1 \cup V_2 = \{v_1,\dots,v_n\}$
%with $v_1 < v_2 < \dots < v_n$ 
and define
\begin{enumerate}[$\bullet$]
\item $Q = \{q_I\} \cup Q_1 \times Q_2 \times \{1,2\} \times (V_1 \cup V_2)$ (where $q_I \not\in Q_1 \cup Q_2$ 
is a new state). Initially, a guess is made of the value $v$ of the input word.
Then, we check that both $A_1$ and $A_2$ visit a weight at least $v$ infinitely often.
In a state $\tuple{q_1,q_2,j,v}$ of $A$, the guess is stored in $v$ (and will never change
along a run) and the value of the index $j$ is toggled to $3-j$ as soon as $A_j$ 
does visit a weight at least $v$;
%\item $I = \{q_I\}$. 
\item For each $\sigma \in \Sigma$, the set $\delta$ contains all the triples 
	\begin{enumerate}[$-$]
	\item $(q_I,\sigma,\tuple{q_1,q_2,1,v})$ such that 
		$v\in V_1 \cup V_2$ and for all $i \in \{1,2\}$, 
		we have $(q_I^{i}, \sigma, q_i) \in \delta_i$.
	\item $(\tuple{q_1,q_2,j,v}, \sigma, \tuple{q'_1,q'_2,j',v'})$ such that
		$v'=v$, $(q_i,\sigma,q'_i) \in \delta_i$ ($i=1,2$), and $j' = 3-j$ if $\weight_j(q_j,\sigma,q'_j) \geq v$,
        	and $j' = j$ otherwise. 
	\end{enumerate}
\item $\weight$ is defined by $\weight(q_I,\sigma,\tuple{q_1,q_2,1,v}) = 0$ and $\weight(\tuple{q_1,q_2,j,v}, \sigma, \tuple{q'_1,q'_2,j',v'})$
is $v$ if $j \neq j'$ and $v_{\min}$ otherwise, where $v_{\min}$ is the minimal weight in $V_1 \cup V_2$.
\end{enumerate}
For the case of $\LimInf$-automata $A_1,A_2$, we can use the synchronized product 
$A_{12} = A_1 \times A_2$, where the weight of a joint transition in $A_{12}$ is the minimum 
of the corresponding transition weights in $A_1$ and $A_2$. It is easy to see that 
$L_{A_{12}} = \min(L_{A_1}, L_{A_2})$ in both cases, and $A_{12}$ is deterministic
when $A_1$ and $A_2$ are deterministic. This case is simpler also because for 
$\LimInf$-automata, deterministic and nondeterministic automata have the same 
expressive power.
\qed

On the negative side, the (deterministic or not) limit-average and discounted-sum 
automata are not closed under $\min$.

\begin{thm}\label{theo:dla-nla-not-closed-under-min}
% \dla\/ and \nla\/ are not closed under the min operator.
The (non)deterministic $\LimAvg$-automata are not closed under $\min$.
\end{thm}

\proof
%The result follows from Lemma~\ref{lem:limavg-min-comp} and the fact that 
%there exists \dla\/ for the languages $L_a$ and $L_b$ of Lemma~\ref{lem:limavg-min-comp}.

Consider the alphabet $\Sigma=\set{a,b}$, and consider the languages
$L_a$ and $L_b$ that assign the long-run average number 
of $a$'s and $b$'s, respectively. Note that there exist \dla\/ for the languages $L_a$ and $L_b$.

We show that there is no \nla\/ for the language $L_m=\min\set{L_a,L_b}$.
To obtain a contradiction, assume that there exists a \nla\/ $A$ for $L_m$. 
We first claim that there must be either an $a$-cycle or a 
$b$-cycle $C$ that is reachable in $A$ such that the sum of the weights in $C$ is positive.
Otherwise, if for all $a$-cycles and $b$-cycles we have that 
the sum of the weights is zero or negative, then we fool the automaton
as follows. Let $\beta$ be the maximum of the absolute values of the 
weights in $A$, and let $\alpha=\lceil \beta \rceil$. 
Then consider the word 
$w=(a^{5\cdot \alpha \cdot \abs{Q}} \cdot b^{5\cdot\alpha \cdot \abs{Q}})^\omega$.
For a run $r$ of $A$ over $w$, the long-run average of the
weights is bounded as follows:
\[
\frac{4 \cdot \beta \cdot \abs{Q}}{10\cdot \alpha\cdot \abs{Q}} \leq \frac{2}{5}.
\]
The above bound is as follows: in the run over $a^{5 \cdot\alpha \cdot \abs{Q}}$, there 
can be a prefix of size at most $\abs{Q}$ with sum of weights at most 
$\abs{Q}\cdot \beta$, and then there would be $a$-cycles, and then a trailing
prefix of size at most $\abs{Q}$ with sum of weights at most $\abs{Q}\cdot \beta$.
Similar argument holds for the segment of $b^{5 \cdot \alpha \cdot \abs{Q}}$.
Hence $L_{A}(w)\leq \frac{2}{5}$, however, $L_m(w)=\frac{1}{2}$,
i.e., we have a contradiction.
W.l.o.g., we assume that there is an $a$-cycle $C$ such that the sum of weights of $C$ 
is positive.
Then we present the following word $w$: a finite word $w_C$ to reach 
the cycle $C$, followed by $a^\omega$; the answer of the automaton is positive, 
{\it i.e.}, $L_{A}(w)>0$,  while $L_m(w)=0$.
Hence the result follows.
\qed

\noindent Finally, we show that discounted-sum automata are not closed under $\min$.

\newcommand{\disc}{\mathit{discount}}

\begin{thm}\label{theo:disc-min}
% \ddi\/ and \nDi\/ are not closed under the min operator.
The (non)deterministic $\Disc$-automata are not closed under $\min$.
\end{thm}

\proof
Let $\lambda$ be a non-algebraic number in $]\frac{1}{2},1[$. 
We consider the quantitative languages 
$L_a^\lambda$ and $L_b^\lambda$ that assign the $\lambda$-discounted sum 
of $a$'s and $b$'s, respectively. 
Formally, given a (finite or infinite) word $w = w_0 w_1 \dots \in \Sigma^* \cup \Sigma^{\omega}$, let
$$\reward_a(w) = \sum_{i\mid w_i = a}^{\abs{w}} \lambda^i \quad \text{ and } \quad 
\reward_b(w) = \sum_{i\mid w_i = b}^{\abs{w}} \lambda^i $$
be the $\lambda$-discounted sum of the $a$'s (resp. $b$'s) of $w$. 
Then, $L_a^\lambda(w) = \reward_a(w)$ and $L_b^\lambda(w) = \reward_b(w)$. 
These languages are definable by \ddi. 
We show that the language $L_m =\min(L_a^\lambda, L_b^\lambda)$ is not definable by a \nDi.

Assume towards contradiction that there is a \nDi\/ $A$ for $L_m$.
%By Lemma~\ref{lem:exist-ambiguous}, there exists an infinite word $\hat{w}$
%all of whose prefixes are ambiguous and thus $\reward_a(\hat{w}) = \reward_b(\hat{w})$ 
%by Lemma~\ref{lem:condition-ambiguous}. 
By Lemmas~2 and~3 in~\cite{CDH08}, there exists an infinite word $\hat{w}$
such that $\reward_a(\hat{w}) = \reward_b(\hat{w})$.

Since $\reward_a(\hat{w}) + \reward_b(\hat{w}) = \frac{1}{1-\lambda}$,
we have $L_m(\hat{w}) = \frac{1}{2(1-\lambda)}$ and this is the maximal
value of a word in $L_m(\cdot)$. 

The maximal value in the automaton $A$ can be obtained for a lasso-word 
of the form $w_1.(w_2)^{\omega}$ (where $w_1,w_2$ are finite words 
and $w_2$ is nonempty), as pure memoryless strategies exist in games 
over finite graphs with the objective to maximize the discounted sum 
of payoffs. Since the language of $A$ is $L_m$, the value of $w_1.(w_2)^{\omega}$
is $\frac{1}{2(1-\lambda)}$, and thus $\reward_a(w_1.(w_2)^{\omega}) = \reward_b(w_1.(w_2)^{\omega})$ by
a similar argument as above. This last condition can be written as
$$p_a(\lambda) + \frac{\lambda^{n_1}\cdot q_a(\lambda)}{1-\lambda^{n_2}} = p_b(\lambda) + \frac{\lambda^{n_1}\cdot q_b(\lambda)}{1-\lambda^{n_2}}$$
for some polynomials $p_a, p_b, q_a, q_b$ and integers $n_1 \geq 0$ and $n_2 > 0$,
or more simply as
\begin{equation}
(1-\lambda^{n_2})\cdot p(\lambda) + \lambda^{n_1}\cdot q(\lambda) = 0 \label{eq:polynomial-lambda}
\end{equation}
for some polynomials $p$ of degree $n_1-1$ and $q$ of degree $n_2-1$,
all of whose coefficients are either $1$ or $-1$. Equation~\eqref{eq:polynomial-lambda}
is not identically zero as either $(i)$ $n_1 = 0$ and it reduces to $q(\lambda) = 0$
or $(ii)$ $n_1 > 0$ and then $p$ has degree at least $0$ so that the term of degree 
zero is not null in~\eqref{eq:polynomial-lambda}. 

Therefore, $\lambda$ must be algebraic, a contradiction.
\qed

\subsection{Closure under complement for infinite words}
% % % % % % % % % % % % % % % % % % % % % % % % % % % % % % % % % % % % % % % % % % % % % % % % % % % % % % % % % % % % 

%The \emph{complement} of a quantitative language $L$ is the 
%quantitative language $L'$ such that $L'(w) = 1-L(w)$ for all
%$w \in \Sigma^{\omega}$.

Most of the weighted automata are not closed under complement.
The next result is a direct extension of the boolean case.

\begin{thm}\label{theo:max-liminf-limsup-not-closed-under-complement}
% \dmax, \nmax, \dls, \dli\/ and \nli\/ are not closed under complement.
The (non)deterministic $\Max$- and $\LimInf$-automata, and the deterministic 
$\LimSup$-automata are not closed under complement.
\end{thm}

\proof
The result follows from a similar result for the boolean version of these classes.
For \dmax\/ and \nmax, consider the language $L_1$ over $\Sigma=\{a,b\}$ 
such that $L_1(a^{\omega}) = 0$ and $L_1(w) = 1$ for all $w \neq a^{\omega}$.
For \dli\/ and \nli, consider the language $L_2$ over $\Sigma=\{a,b\}$ 
such that $L_2(\Sigma^*.a^{\omega}) = 1$ and $L(w) = 0$ for all words $w$ containing 
infinitely many $b$'s, and for \dls, consider $L_3$ the complement of $L_2$.
\qed

The next theorem is a positive result of closure under complementation
for \nls. It reduces to the complementation of nondeterministic B\"uchi automata.

\begin{thm}\label{theo:nls-closed-under-complement}
% \nls\/ is closed under complement.
The nondeterministic $\LimSup$-automata are closed under complement, with cost $O(m \cdot 2^{n \log n})$.
\end{thm}

\proof
Let $A=\tuple{Q,q_{0},\Sigma,\delta,\weight}$ be a \nls,
and let $V = \{\weight(e) \mid e \in \delta\}$
be the set of weights that appear in $A$.
For each $v \in V$, it is easy to construct a \nbw\/ $A_v$
whose (boolean) language is the set of words $w$
such that $L_A(w) \geq v$, by declaring to be accepting
the edges with weight at least $v$.
We then construct for each $v \in V$ a \nbw\/ $\bar{A}_v$ (with accepting edges) 
that accepts the (boolean) complement of the language accepted by $A_v$.
Finally, assuming that $V= \{v_1,\dots,v_n\}$ with $v_1 < v_2 < \dots < v_n$,
we construct the \nls\/ $B_i$ for $i=2,\dots,n$ where $B_i$ is obtained
from $\bar{A}_{v_i}$ by assigning weight $1-v_{i-1}$ to each accepting edges,
and $1-v_n$ to all other edges. The complement of $L_A$ is then
$\max\{L_{B_2},\dots,L_{B_n}\}$ which is accepted by a \nls\/ by Theorem~\ref{theo:max-closure}.
\qed

\begin{thm}
% \ddi\/ is closed under complement.
The deterministic $\Disc$-automata are closed under complement, with cost $O(n)$.
\end{thm}

\proof[sketch]
It suffices to replace each weight $v$ of a \ddi\/ by $1-\lambda-v$
(where $\lambda$ is the discount factor) to obtain the \ddi\/ for
the complement. 
\qed

\begin{thm}
% \dla\/ is not closed under complement.
The deterministic $\LimAvg$-automata are not closed under complement.
\end{thm}

\begin{figure}[t]
   \begin{center}
  \begin{tikzpicture}[node distance=1.8cm,auto,shorten >=1 pt,>=latex']
  \node[rdbox,initial,initial text={}]  (0)  at (0,0) {} ;
  \draw[->] (0) edge[loop,out=120,in=60,looseness=8] 
                node[above]       {$a,1$}(0);
%                node[below]       {0}(0);
  \draw[->] (0) edge[loop,out=240, in=300,looseness=8] 
                node[below]       {$b,0$}(0);
%                node[below]       {2}(1);
  \end{tikzpicture}
   \end{center}
  \caption{Deterministic Limit-average Automaton.}
  \label{figure:aut1}
\end{figure}
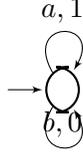 

\proof
Consider the \dla\/ $A$ over alphabet $\Sigma= \{a,b\}$ 
(shown in Figure~\ref{figure:aut1}) that consists of 
a single self-loop state with weight $1$ for $a$ and $0$ for $b$.
Notice that $L_A(w.a^\omega) = 1$ and $L_A(w.b^\omega) = 0$ for all $w \in \Sigma^*$.
To obtain a contradiction, assume that there exists a \dla\/ $B$ 
whose language is $L_B = 1-L_A$. For all finite words $w \in \Sigma^*$, 
let $L^{\Avg}_B(w)$ be the average weight of the unique (finite) run of $B$ over $w$.

Fix $0 < \epsilon < \frac{1}{2}$. For all finite words $w$, there exists
a number $n_w$ such that the average number of $a$'s in $w.b^{n_w}$ is at most $\epsilon$, 
and there exists a number $m_w$ such that $L^{\Avg}_B(w.a^{m_w}) \leq \epsilon$
(since $L_B(w.a^\omega) = 0$). Hence, we can construct a word 
$w = b^{n_1} a^{m_1} b^{n_2} a^{m_2} \dots$ such that 
$L_A(w) \leq \epsilon$ and $L_B(w) \leq \epsilon$.
Since $L_B = 1-L_A$, this implies that $1 \leq 2\epsilon$, a contradiction.
\qed

\begin{thm}
% \nla, and \nDi\/ are not closed under complement.
The nondeterministic $\LimAvg$- and $\Disc$-automata are not closed under complement.
\end{thm}

\proof
The fact that \nla\/ are not closed under complementation is
as follows. Consider the quantitative language $L^*=1-\max\set{L_a,L_b}$
where $L_a$ and $L_b$  assign the long-run average number of $a$'s and $b$'s, respectively.
Exactly the same argument as in the proof of Theorem~\ref{theo:dla-nla-not-closed-under-min}
shows that $L^*$ cannot be expressed as a 
\nla, while the language $\max\set{L_a,L_b}$ can be expressed
as \nla\/ by Theorem~\ref{theo:max-closure}.

That \nDi\/  are not closed under complement can be 
obtained as follows: given $0<\lambda<1$, consider the 
language $L_a^\lambda$ and $L_b^\lambda$ that assigns to words the
$\lambda$-discounted sum of $a$'s and $b$'s, respectively.
The language $L_a^\lambda$ and $L_b^\lambda$ can be expressed
as \ddi, and the max of them can be defined by \nDi.
Observe that $L_a^\lambda(w)+L_b^\lambda(w) = \frac{1}{1-\lambda}$ 
for all $w \in \Sigma^{\omega}$. Therefore, 
$\min\set{L_a^\lambda,L_b^\lambda}=\frac{1}{1-\lambda}-\max\set{L_a^\lambda,L_b^\lambda}$.
Since \nDi\/ is not closed under min (Theorem~\ref{theo:disc-min}), we 
immediately obtain that \nDi\/ are not closed under complementation.
\qed

% % % % % % % % % % % % % % % % % % % % % % % % % % % % % % % % % % % % % % % % % % % % % % % % % % % % % % % % % % % % 
\subsection{Closure under sum for infinite words}
% % % % % % % % % % % % % % % % % % % % % % % % % % % % % % % % % % % % % % % % % % % % % % % % % % % % % % % % % % % % 

All weighted automata are closed under sum,
except \dla\/ and \nla.

\begin{thm}\label{theo:max-closed-under-sum}
% \dmax\/ and \nmax\/ are closed under sum, with cost $O(n_1\cdot m_1 \cdot n_2 \cdot m_2)$.
The (non)deterministic $\Max$-automata are closed under sum, with cost $O(n_1\cdot m_1 \cdot n_2 \cdot m_2)$.
\end{thm}

\proof
The construction is the same as for $\Maxf$-automata over finite words
given in the proof of Theorem~\ref{theo:closure-under-sum-finite}.
\qed

\begin{thm}\label{theo:nls-closed-under-sum}
% \nls\/ is closed under sum, with cost $O(n_1\cdot m_1 \cdot n_2 \cdot m_2)$.
The nondeterministic $\LimSup$-automata are closed under sum, with cost $O(n_1\cdot m_1 \cdot n_2 \cdot m_2)$.
\end{thm}

\proof[Sketch]
Given two \nls\/ $A_1$ and $A_2$, we construct a \nls\/ $A$
for the sum of their languages as follows. Initially, we make a guess of
a pair $(v_1,v_2)$ of weights ($v_i$ in $A_i$, for $i=1,2$) and we branch
to a copy of the synchronized product of $A_1$ and $A_2$. We attach a bit $b$
whose range is $\{1,2\}$ to each state to remember that we expect $A_b$ to 
visit the guessed weight $v_b$. Whenever this occurs, the bit $b$ is set to $3-b$,
and the weight of the transition is $v_1 + v_2$. All other transitions ({\it i.e.}
when $b$ is unchanged) have weight $\min\{v_1 + v_2 \mid v_1 \in V_1 \land v_2 \in V_2\}$.
\qed

\begin{thm}
% \dls\/ is closed under sum, with cost $O(n_1\cdot n_2 \cdot 2^{m_1 \cdot m_2})$.
The deterministic $\LimSup$-automata are closed under sum, with cost $O(n_1\cdot n_2 \cdot 2^{m_1 \cdot m_2})$.
\end{thm}

\proof
Let $A_1=\tuple{Q_1,q_I^1,\Sigma,\delta_1,\weight_1}$ and $A_2=\tuple{Q_2,q_I^2,\Sigma,\delta_2,\weight_2}$
be two \dls. We construct a \dls\/ $A=\tuple{Q,q_I,\Sigma,\delta,\weight}$ such that
$L_A = L_{A_1} + L_{A_2}$. Let $V_i = \{\weight_i(e) \mid e \in \delta_i\}$
be the set of weights that appear in $A_i$ (for $i=1,2$). 
The automaton $A$ implements the synchronized product of $A_1$ and $A_2$,
and keeps one bit $b(v_1,v_2)$ for each pair $(v_1,v_2)$ of weights $v_1 \in V_1$ and $v_2 \in V_2$.
For $i=1,2$, if $b(v_1,v_2)=i$, then $A_i$ is expected to cross a transition with weight
$v_i$. Whenever this occurs, the bit is set to $3-i$.
The weight of a transition in $A$ is the largest value of $v_1 + v_2$
such that the corresponding bit $b(v_1,v_2)$ has changed in the transition.
Formally, we define:
\begin{enumerate}[$\bullet$]
\item $Q = Q_1 \times Q_2 \times [V_1 \times V_2 \to \{1,2\}]$;
\item $q_I = \tuple{q_I^1,q_I^2,b_I}$ where $b_I(v_1,v_2) = 1$ for all $(v_1,v_2) \in V_1 \times V_2$;
\item For each $\sigma \in \Sigma$, the set $\delta$ contains all the triples 
 $(\tuple{q_1,q_2,b}, \sigma, \tuple{q'_1,q'_2,b'})$ such that
	$(q_i,\sigma,q'_i) \in \delta_i$ ($i=1,2$), and for all $(v_1,v_2) \in V_1 \times V_2$, 
	we have $b'(v_1,v_2) = 3-b(v_1,v_2)$ 
	if $\weight_i(\tuple{q_i,\sigma,q'_i}) = v_i$ for $i=b(v_1,v_2)$,
	and otherwise $b'(v_1,v_2) = b(v_1,v_2)$.
\item $\weight$ is defined by $\weight(\tuple{q_1,q_2,b}, \sigma, \tuple{q'_1,q'_2,b'}) = 
\max(\{v_{\min} \cup \{v_1 + v_2 \mid b'(v_1,v_2) \neq b(v_1,v_2) \})$
where $v_{\min}$ is the minimal weight in $V_1 + V_2 = \{v_1 + v_2 \mid v_1 \in V_1 \land v_2 \in V_2\}$.\qed
\end{enumerate}

\begin{thm}
% \dli\/ and \nli\/ are closed under sum, with cost $O(n_1\cdot n_2 \cdot 2^{m_1 \cdot m_2})$.
The (non)deterministic $\LimInf$-automata are closed under sum with cost $O(n_1\cdot n_2 \cdot 2^{m_1 \cdot m_2})$.
\end{thm}

\proof
Let $A_1=\tuple{Q_1,q_I^1,\Sigma,\delta_1,\weight_1}$ and $A_2=\tuple{Q_2,q_I^2,\Sigma,\delta_2,\weight_2}$
be two \nli. We construct a \nli\/ $A=\tuple{Q,q_I,\Sigma,\delta,\weight}$ such that
$L_A = L_{A_1} + L_{A_2}$. Let $V_i = \{\weight_i(e) \mid e \in \delta_i\}$
be the set of weights that appear in $A_i$ (for $i=1,2$). 
The automaton $A$ implements the synchronized product of $A_1$ and $A_2$,
and keeps one bit $b(v_1,v_2)$ for each pair $(v_1,v_2)$ of weights $v_1 \in V_1$ and $v_2 \in V_2$.
If a transition in $A_i$ for some $i \in \{1,2\}$ has weight less than $v_i$,
then the bit $b(v_1,v_2)$ is set to $\bot$, otherwise is set to $\top$.
The weight of a transition in $A$ is the largest value of $v_1 + v_2$
such that the corresponding bit $b(v_1,v_2)$ is $\top$.
Formally, we define:
\begin{enumerate}[$\bullet$]
\item $Q = Q_1 \times Q_2 \times [V_1 \times V_2 \to \{\top,\bot\}]$;
\item $q_I = \tuple{q_I^1,q_I^2,b_I}$ where $b_I(v_1,v_2) = \bot$ for all $(v_1,v_2) \in V_1 \times V_2$;
\item For each $\sigma \in \Sigma$, the set $\delta$ contains all the triples 
 $(\tuple{q_1,q_2,b}, \sigma, \tuple{q'_1,q'_2,b'})$ such that
	$(q_i,\sigma,q'_i) \in \delta_i$ ($i=1,2$), and for all $(v_1,v_2) \in V_1 \times V_2$, 
	we have $b'(v_1,v_2) = \top$ 
	if $\weight_i(\tuple{q_i,\sigma,q'_i}) \geq v_i$ for $i=1,2$,
	and otherwise $b'(v_1,v_2) = \bot$.
\item $\weight$ is defined by $\weight(\tuple{q_1,q_2,b}, \sigma, \tuple{q'_1,q'_2,b'}) = 
\max(\{v_{\min} \cup \{v_1 + v_2 \mid b'(v_1,v_2) = \top \})$
where $v_{\min}$ is the minimal weight in $V_1 + V_2 = \{v_1 + v_2 \mid v_1 \in V_1 \land v_2 \in V_2\}$.
\end{enumerate}
The result for \dli\/ follows from the fact $A$ is deterministic if
$A_1$ and $A_2$ are deterministic.
\qed

\begin{thm}\label{theo:ddi-ndi-closed-under-sum}
% \ddi\/ and \nDi\/ are closed under sum, with cost $O(n_1 \cdot n_2)$.
The (non)deterministic $\Disc$-automata are closed under sum, with cost $O(n_1 \cdot n_2)$.
\end{thm}

\proof[Sketch]
It is easy to see that the synchronized product of two \nDi\/ (resp. \ddi)
defines the sum of their languages, if the weight of a joint transition
is defined as the sum of the weights of the corresponding transitions in the two \nDi\/ (resp. \ddi).
\qed

\begin{thm}
% \dla\/ and \nla\/ are not closed under sum.
The (non)deterministic $\LimAvg$-automata are not closed under sum.
\end{thm}

\proof
Consider the alphabet $\Sigma=\set{a,b}$, and consider the \dla-definable languages 
$L_a$ and $L_b$ that assigns to each word $w$ the long-run average number of $a$'s and $b$'s in $w$
respectively. 
Let $L_{+}=L_a + L_b$. Assume that $L_{+}$ is defined by a \nla\/ $A$ with set 
of states $Q$ (we assume w.l.o.g that every state in $Q$ is reachable).

First, we claim that from every state $q \in Q$, there is a run of $A$ over $a^{\abs{Q}}$
that visit a cycle $C^*$ with average weight $1$. To see this, notice that from every state $q \in Q$,
there is an infinite run $\rho$ of $A$ over $a^\omega$ whose value is $1$ 
(since $L_{+}(w_q\cdot a^\omega) =1$ for all finite words $w_q$). 
Consider the following decomposition of $\rho$.
Starting with an empty stack, we push the states of $\rho$ onto the stack
as soon as all the states on the stack are different. If the next state is already
on the stack, we pop all the states down to the repeated state thus removing a simple
cycle of $\rho$. Let $C_1$, $C_2, \dots$ be the cycles that are successively 
removed. Observe that the height of the stack is always at most $\abs{Q}$.
Let $\beta$ be the largest average weight of the cycles $C_i$, $i\geq 1$,
and let $\alpha_{\max}$ be the largest weight in $A$.
Assume towards contradiction that $\beta < 1$. Then, for all $n > 0$, 
the value of the prefix of length $n$ of $\rho$ is at most:
$$ \frac{\alpha_{\max} \cdot \abs{Q} + \beta \cdot \sum_{i=1}^{k_n} \abs{C_i}}{n} $$
where $k_n$ is the number of cycles that have been removed from the stack 
when reading the first $n$ symbols of $\rho$. Hence, the value of $\rho$ is at most $\beta < 1$,
which is a contradiction. Therefore, the average weight of some cycle $C^* = C_i$ 
is exactly\footnote{It cannot be greater than $1$ since $L_{+}(w\cdot a^\omega)=1$ for all finite 
words $w$.} $1$ (there are finitely many different cycles as they are simple cycles).
Since the height of the stack is at most $\abs{Q}$, the cycle $C^*$ is reachable
in at most $\abs{Q}$ steps.

Second, it can be shown analogously that from every state $q \in Q$, there is a 
run over $b^{\abs{Q}}$ that visit a cycle $C^*$ with average weight $1$.

Third, for arbitrarily small $\epsilon >0$, consider the word $w$ and the run $\rho$ of $A$ over $w$ generated inductively 
by the following procedure: $w_0$ is the empty word and $\rho_0$ is the initial state of $A$
We generate $w_{i+1}$ and $\rho_{i+1}$ from $w_i$ and $\rho_i$ as follows:
\begin{compressEnum}
\itCompress generate a long enough sequence $w_{i+1}'$ of $a$'s after $w_i$ such that the 
average number of $b$'s in $w_i \cdot w_{i+1}'$ falls below $\epsilon$
and we can continue $\rho_i$ and reach within at most $\abs{Q}$ steps (and then repeat $k$ times) 
a cycle $C$ of average weight $1$ and such that the average weight of this run
prolonged by $\abs{Q}$ arbitrary transitions is at least $1-\epsilon$, {\it  i.e.}
$$ \frac{\gamma(\rho_i)+k\cdot\abs{C} + 2 \alpha_{\min} \cdot \abs{Q} }{\abs{\rho_i} + k\cdot \abs{C} + 2 \cdot \abs{Q}}
\geq 1-\epsilon$$
where $\alpha_{\min}$ is the least weight in $A$. This is possible since $k$ can be chosen
arbitrarily large. Let $\rho'_i$ be the prolongation of $\rho_i$ over $w_{i+1}'$;
\itCompress then generate a long enough sequence $w_{i+1}''$ of $b$'s such that the average number of 
$a$'s in $w_{i} \cdot w_{i+1}' \cdot w_{i+1}''$ falls below $\epsilon$
and as above, we can construct a continuation $\rho''_i$ of $\rho'_i$ whose
average weight is at least $1-\epsilon$ (even if prolonged by $\abs{Q}$ arbitrary transitions);
\itCompress the word $w_{i+1}=w_i \cdot w_{i+1}' \cdot w_{i+1}''$ and the run 
$\rho_{i+1}$ is $\rho''_i$.
\end{compressEnum}
The word $w$ and the run $\rho$ are the limit of these sequences.
We have $L_{a}(w) = L_{b}(w) = 0$ and thus $L_{+}(w) = 0$, while
the value of $\rho$ is at least $1-\epsilon$, a contradiction.\qed

\paragraph{{\bf Acknowledgment.}} We thank Wolfgang Thomas for pointing out the isolated cut-point problem.

\bibliography{biblio}
\bibliographystyle{alpha}
\end{document}